\documentclass[12pt,a4paper]{article}

\usepackage{jheppub}

\usepackage{etoolbox}
    \makeatletter
    \patchcmd{\maketitle}{\@fpheader}{}{}{}
    \makeatother

\usepackage{amssymb}
\usepackage{mathtools}

\usepackage{relsize,exscale}
\usepackage{tensor}

\usepackage[utf8]{inputenc}
\usepackage[T1]{fontenc}
\usepackage{lmodern}

\usepackage{mathrsfs}

\usepackage{graphicx}
\usepackage[]{subdepth}

\newcommand{\scr}{\scriptscriptstyle}

\newcommand{\Lint}{\mathop{\mathlarger{\int}} \! \! }

\usepackage{wrapfig}
\usepackage{array}

\usepackage[numbers]{natbib}

\usepackage{pgfplots}
\usetikzlibrary{decorations.pathmorphing}
\usetikzlibrary{arrows}
\usetikzlibrary{patterns}
\usetikzlibrary{snakes}

\title{One-loop electromagnetic correlators of SQED in power-law inflation}

\author[1]{Dra\v{z}en Glavan,}
\emailAdd{drazen.glavan@uclouvain.be}
\author[2]{Gerasimos Rigopoulos}
\emailAdd{gerasimos.rigopoulos@newcastle.ac.uk}

\affiliation[1]{
	Centre for Cosmology, Particle Physics and Phenomenology (CP3),
	\\
	Universit\'{e} catholique de Louvain, 
	\\
	Chemin du Cyclotron 2, 1348 Louvain-la-Neuve, Belgium
	}
\affiliation[2]{
	School of Mathematics, Statistics and Physics,
	\\
	Herschel Building, Newcastle University,
	\\
	Newcastle upon Tyne, NE1 7RU, United Kingdom}

\abstract{

We examine scalar quantum electrodynamics in power-law inflation,
and compute the one-loop correction to electric and magnetic field
correlators at superhorizon separations. The effect at one-loop descends
from the coupling of the vector to the charged scalar current which is 
greatly enhanced due to gravitational particle production.
We conclude that non-perturbative effects must exist due to (i) secular growth,
(ii) spatial running, and (iii) infrared sensitivity of the one-loop correction
to the correlators. Electric and magnetic correlators exhibit a hierarchy that
is due to Faraday's law and accelerated expansion,
and must hold non-perturbatively.
}

\preprint{{\tt CP3-19-44}}

\begin{document}

\notoc
\maketitle

\vspace{3cm}

\section{Introduction}
\label{sec: Introduction}

The idea of utilizing the primordial inflationary universe as a testing ground
for Beyond Standard Model (BSM) particle physics -- dubbed the {\it Cosmological Collider} -- \
has gained prominence lately~\cite{Arkani-Hamed:2015bza,Meerburg:2016zdz}. 
The high energy scale of the expansion in primordial inflation can excite very heavy matter fields 
inaccessible to foreseeable Earth-base accelerators. These, in turn, could have left an imprint
on cosmological observables.
However, if we are to sift through the signals from primordial inflation for signs of BSM physics, 
it is paramount that we first understand the behaviour of the Standard Model (SM) itself in
the extreme gravitational conditions of the rapidly expanding and accelerating Universe.

\begin{wrapfigure}{l}{0.6\textwidth}
\centering
\medskip
\includegraphics[width=9.cm]{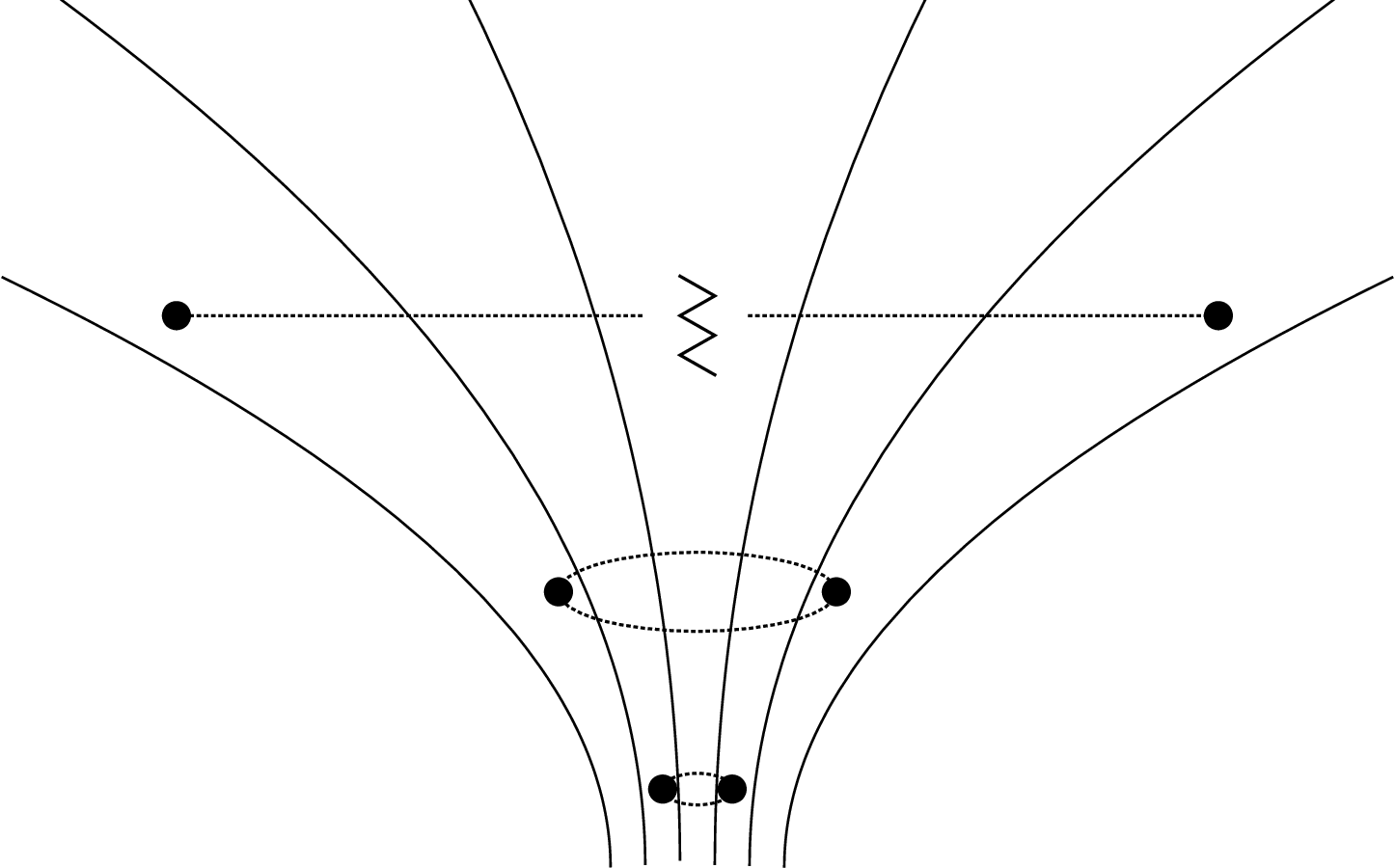}
\caption{\linespread{1}\selectfont 
Gravitational particle production. A virtual pair created as a quantum
fluctuation is stretch rapidly by the expanding space to the point 
where its lifetime becomes infinite, and it is ripped out of the vacuum into
existence.}
\vskip-0.1cm
\label{grav part prod}
\end{wrapfigure}

The SM fields are all light in inflation, relative to the
Hubble scale, $H\!\sim\!10^{13}$GeV, which can lead to drastically
different phenomenology compared to flat space.
In particular, the fields that couple non-conformally to gravity --  light
scalars and the graviton -- are influenced by the expansion the most.
Even at tree-level they experience the so-called gravitational
particle production~\cite{Parker:1968mv,Parker:1969au,Birrell:1982ix,Mukhanov:2007zz,Parker:2009uva}. 
The virtual pairs of particles, popping in and out of 
existence due to quantum fluctuations, are stretched to superhorizon 
separations by the rapid expansion before they can annihilate,
and thus the expansion rips them from the vacuum into existence,
as depicted schematically in Fig.~\ref{grav part prod}.
This infrared (IR) effect is so strong that in general it leads to the symmetry restoration
of symmetry breaking potentials in scalar 
models~\cite{Ford:1985qh,Starobinsky:1994bd,Lazzari:2013boa,Serreau:2013eoa,Guilleux:2015pma}.
The Higgs field of the SM is extremely light in inflation, so it experiences
large IR effects.


On the other hand, fields that couple conformally to gravity, 
such as gauge bosons or fermions, do not experience 
gravitational particle productions in conformally flat cosmological
space-times, as they effectively do not distinguish between
the expanding space and flat space.~\footnote{\linespread{1}\selectfont 
Quantum effects generally make the conformal symmetry 
anomalous~\cite{Dolgov:1981nw}, which breaks
conformality only mildly (see also~\cite{Benevides:2018mwx}
for the implications on primordial magnetogenesis).}
Nevertheless, couplings of those fields to the graviton, or a light scalar
such as the SM Higgs field, can communicate to them  the effects of the expansion.
Therefore, it is to be expected that all SM fields can be influenced
by the expansion of primordial inflation, and it is important
to understand what kind of signals they could impart on the cosmological observables.
This question was emphasized recently 
in~\cite{Chen:2016nrs,Chen:2016uwp,Chen:2016hrz}, where the authors 
concentrated on how inflation changes the mass spectrum of the SM. 
However, as we show here, mass corrections might not be the only signal of the SM in inflation. 

\

\medskip

\begin{wrapfigure}{r}{0.6\textwidth}
\centering
\includegraphics[width=9.cm]{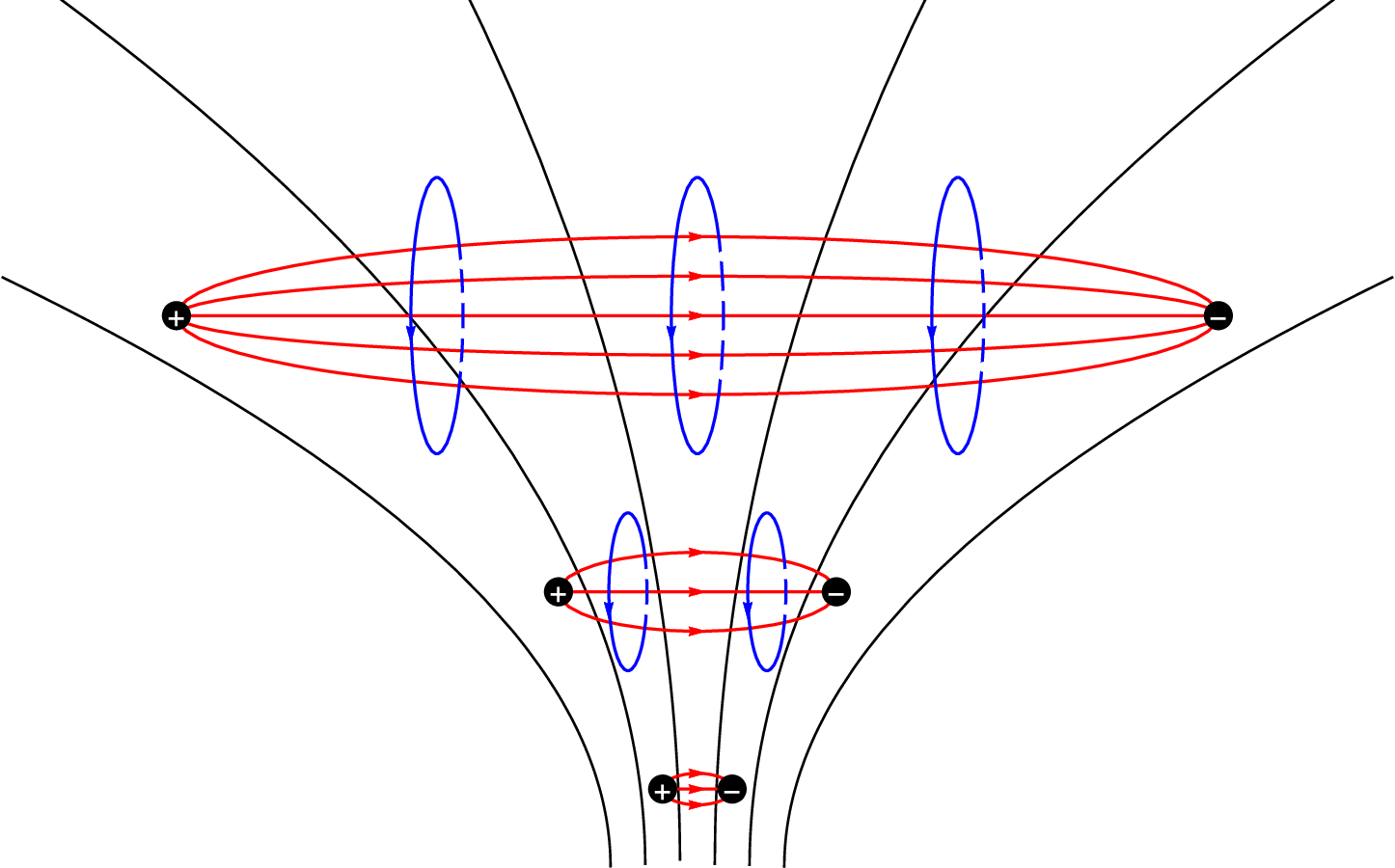}
\caption{\linespread{1}\selectfont 
Gravitational particle production of charged particles.
Particles of the pair created by gravitational particle production
carry opposite charges, and couple to the photon field. The photon field
reacts to created charges resulting in electric (red) and magnetic (blue) fields.
The created electric and magnetic fields further act on the charges,
and may lead to further creation of charged pairs via Schwinger mechanism.}
\label{charged part prod}
\end{wrapfigure}

This work is primarily motivated by the question of how do the boson
fields of the SM electroweak (EW) sector behave
in primordial inflation, and in particular how are they affected by
loop corrections from the
Higgs field with large IR enhancement. To this end we study a simplified model -- 
scalar quantum electrodynamics (SQED) -- that contains a light complex scalar
with~$U(1)$ charge, and a vector that couples to it. Since the complex scalar
 is charged, gravitational particle production 
now creates pairs of opposite charge. The vector field couples to the created charges
and the currents they produce, establishing electric and magnetic (E\&M) fields.
As the charged pair is stretched to superhorizon separations, so are the 
E\&M fields, as is schematically depicted in Fig.~\ref{charged part prod}.

The reaction of the vector field to gravitationally induced
charged currents, and the resulting
superhorizon electric and magnetic fields, are one-loop quantum effects.
In this work we compute the one-loop correction to the 
superhorizon correlators
of quantum fluctutions of E\&M fields,
\begin{equation}
\bigl\langle \hat{F}_{\mu\nu}(\eta,\vec{x}) \, \hat{F}_{\rho\sigma}(\eta,\vec{x}^{\,\prime}) \bigr\rangle \, ,
\end{equation}
for SQED in {\it power-law inflation}.The equal-time E\&M field correlators, induced here due to
interacting quantum fluctuations, bear observational relevance in relation to
the problem of cosmic magnetic fields,~\footnote{
See~\cite{Maleknejad:2012fw,Durrer:2013pga,Subramanian:2015lua} for recent reviews.}
and their primordial origin~\cite{Turner:1987bw,Giovannini:2000dj,Calzetta:1997ku, Kandus:1999st}.
The power-law inflation, characterized by a constant non-zero slow-roll 
parameter,~$\epsilon\!=\!-\dot{H}/H^2$, provides a better approximation for the
more realistic slow-roll inflation,
compared to the de Sitter space,~$\epsilon\!=\!0$, in which most previous works are set.
This is crucial for secular effects coming from quantum loops, 
whose scaling generally depends on~$\epsilon$.

\medskip


Investigations of the quantum loop effects of SQED in inflation 
originated in~\cite{Davis:2000zp,Dimopoulos:2001wx},
where it was argued that the vector field develops an effective mass due to the
large IR fluctuations of the complex scalar. This argument was based on the
Hartree approximation descending from the 4-point interaction, 
which resembles a photon mass. It was inferred that
the generation of such a mass has potential observational significance in 
creating cosmological magnetic fields.
Even though the Hartree approximation is known
not to be reliable quantitatively, the conjecture of the photon mass
generation in massless SQED in de Sitter 
was nevertheless subsequently confirmed 
for a photon propagating through an ensemble of superhorizon quantum fluctuations
in~\cite{Prokopec:2002uw,Prokopec:2002jn,Prokopec:2003bx},
where the full one-loop computation of the vacuum polarization was performed.
However, unlike the original suggestion of~\cite{Davis:2000zp,Dimopoulos:2001wx}, the origin of the photon mass turned out not to be the 4-point interaction, but rather the 3-point one.~\footnote{\linespread{1}\selectfont 
The contribution of the
4-point interaction to the vacuum polarization is precisely canceled by the 
local part of the 3-point contribution. The only local term surviving is the conformal 
anomaly.} The mass of the photon was inferred by examining the one-loop correction
to the photon mode function,
and it was found that the sub-Hubble dynamics is captured by a Proca-like photon mass
term,~$m_\gamma^2 \!=\! \frac{(qH)^2}{2\pi^2}\!\times\!\ln(k/H)$,
where~$k$ is the modulus of the photon's wave-vector, and~$H$ is the de Sitter
space Hubble rate, and~$q$ is the charge of the scalar. 
Furthermore, in~\cite{Prokopec:2003iu} the one-loop corrected
equations for electric and magnetic fields were solved self-consistently,
and it was reported that the electric field of the long-wavelength photon receives power-law secular
enhancement~$\sim a^{1/2}$, while the magnetic field gets a constant 
amplitude renormalization.
The conclusion was that the dynamics of the photon was more intricate than
it just developing a local effective mass.
The investigation was extended to light massive
and non-minimally coupled SQED in~\cite{Prokopec:2003tm,Prokopec:2004au},
with similar conclusions, the generated photon mass 
being~$m_\gamma^2 \!=\! \frac{(qH)^2}{2\pi^2}\!\times\! 3H^2/[2(m_\phi^2\!+\!12\xi H^2)]$,
where~$m_\phi$ is the complex scalar's mass, and~$\xi$ is the dimensionless non-minimal
coupling (see also~\cite{Popov:2017xut}). Again, the effect in the massive case also 
derives only from the
3-point interaction.

Investigations of SQED in de Sitter were greatly
advanced in the seminal work~\cite{Prokopec:2007ak}, where the authors
adapted Starobinsky's stochastic formalism~\cite{Starobinsky:1986fx} 
to SQED in de Sitter.
This was accomplished in two steps: (i) the vector field was integrated out exactly,
owing to it appearing only quadratically in the action, and the effective non-local 
scalar theory was obtained, and (ii) Starobinsky's stochastic equation was derived 
for the gauge-invariant scalar field variance~$\Phi^*\Phi$ from the effective scalar theory,
and solved in the late-time limit. Thus, the leading order IR effects on the scalar have been 
resummed~\footnote{\linespread{1}\selectfont 
For one-loop corrections the photon
induces on the dynamics of the scalar see~\cite{Kahya:2005kj,Kahya:2006ui}.}
to reveal that it develops a non-Gaussian interacting state with a non-perturbatively
large variance,~$\bigl\langle \Phi^* \Phi \bigr\rangle\!\approx\!1.65 \!\times\! H^2\!/q^2$,
but a perturbatively small mass~$m_\phi^2 \!\approx\! 0.45 \!\times\! (qH)^2/(2\pi)^2$. 
This implies that a photon propagating through such a polarizable medium
develops a non-perturbatively large effective 
mass,~$m_\gamma^2 \!=\! 2 q^2 \bigl\langle \Phi^*\Phi \bigr\rangle \!\approx\! 3.3\!\times\! H^2$.
The work on the stochastic formalism was followed up by two-loop computations
of the local scalar bi-linears, and the energy-momentum tensor, 
confirming its consistency~\cite{Prokopec:2006ue,Prokopec:2008gw}.

Recently, comprehensive computations of the one-loop corrections to two-point functions 
of SM fields in de Sitter were reported in~\cite{Chen:2016nrs,Chen:2016uwp,Chen:2016hrz}.
Out of the many relevant cases that the authors considered, here we comment only
on those pertinent to our work -- loop corrections to the 2-point function of
vector bosons in SQED and the EW sector of the SM.
Two diagrams contribute at one-loop to the vector boson propagator correction, 
a local one constructed from a 4-vertex, and a non-local one made up of 3-vertices, 
and both contributions are reported
to exhibit secular growth, which is in accord with other perturbative results.
The correction descending from the local diagram is recognized to have the form of a 
mass insertion, 
which in~\cite{Chen:2016nrs,Chen:2016uwp,Chen:2016hrz} is  resummed using the ansatz of the
Dynamical Renormalization Group (DRG)~\cite{Chen:1995ena,Burgess:2009bs}.
Thus, the authors conclude that the effect the expanding space-time and the coupling to 
a light scalar have on vector fields is to change their mass spectrum. However, it should
be noted that the contribution of the local one-loop diagram cancels out when all one-loop
corrections are accounted for~\cite{Prokopec:2002uw,Prokopec:2003tm}, 
and that it is actually the \emph{non-local diagram}, constructed out of 3-vertices, that provides the leading correction. Furthermore, the DRG is known to treat the
local diagrams correctly, reproducing the known results for the scalar 
fields~\cite{Tsamis:2005hd,Woodard:2005cw,Garbrecht:2013coa,Garbrecht:2014dca,Nacir:2016fzi},
but not for the non-local diagrams relevant for vector fields. Even though, interestingly,
the reported one-loop generated mass~\cite{Chen:2016nrs} of the vector in SQED
matches the one from~\cite{Prokopec:2003tm}, these have a completely different origin.
Moreover, adopting the unitary gauge to infer the loop corrections to the EW sector of
the SM in the symmetric phase seems problematic, since the same arguments applied to SQED 
would eliminate the 3-vertices that are responsible for the dominant one-loop contribution.

\medskip

The computation of the one-loop correction to the equal-time E\&M correlators
that we present here differs from the previous works in the following aspects:
\begin{itemize}
\item
We compute the full two-point function of the E\&M fields in order to quantify the
loop correction, instead of the vector field mode function computed 
in~\cite{Prokopec:2002uw,Prokopec:2008gw}. While it is true that the 
one-loop corrected mode function contains the same information as the
one-loop corrected two-point function, and one can be reconstructed from 
another, they have different physical meaning.
The photon mode function gives the information about the photon that propagates 
through a polarizable medium, the correlators of E\&M fields provide information about
the polarizable medium itself. This is the kind of information closer to the
cosmological observables such as the cosmic microwave background that comes in the
form of correlators across the sky.

\item
We compute a two-point function of a gauge-invariant observable, rather than
the gauge-dependent two-point function of the vector field
as in~\cite{Chen:2016nrs,Chen:2016uwp,Chen:2016hrz}. While there is nothing wrong
with the latter, the physical interpretation is more obscure due to gauge dependence.
This is particularly delicate when it comes to recognizing generated {\it dynamical}
effective mass
of a vector field, which cannot readily be inferred by comparing it to the Proca field.

\item
Effects at superhorizon separations that we consider descend from the
non-local one-loop diagram composed out of 3-vertices of SQED, unlike 
Refs.~\cite{Chen:2016nrs,Chen:2016uwp,Chen:2016hrz} that quote the 
local one-loop diagram, containing only one 4-vertex, as the origin of the photon
mass generation. Our work is in line with Refs.~\cite{Prokopec:2002uw,Prokopec:2003tm}
which demonstrated that the only local effect that does not cancel is 
in fact just the conformal anomaly.

\item
We consider the background to be {\it power-law inflation}, characterized by a finite and constant
slow-roll parameter~$0\!<\!\epsilon\!<\!1$, as opposed to the 
exact de Sitter space-time ($\epsilon\!=\!0$) considered previously.
Observations suggest that primordial inflation had a finite and slowly-changing 
slow-roll parameter of the order~$\epsilon\!\sim\!0.01$, 
as measured by the spectral tilt of primordial fluctuations. 
Power-law inflation does not represent this faithfully, but it does capture the effects
of the finite slow-roll parameter.
Even though the de Sitter approximation is often good, but for
the purposes of loop corrections in inflation it does not have to be so, 
as they typically exhibit secular corrections. A secular correction
that scales as~$\sim \!\ln(a)$ in de Sitter might  exhibit scaling
of the form~$\sim\! (a^{\epsilon}\!-\!1)/\epsilon$ in power-law inflation.
It is not really known how long inflation might have lasted, and given
enough time these two scalings start deviating considerably. This expectation is
corroborated by the computation presented here.
Moreover, working in power-law inflation as opposed to exact de Sitter allows us
to break the degeneracy between the (effective) particle mass and the 
Hubble rate. Some works would suggest that fields tend to develop effective
non-minimal coupling to the Ricci scalar, as opposed to the constant mass 
term~\cite{Janssen:2009pb,Glavan:2020zne}.
The two have different scalings, and it is important to distinguish between the two.

\end{itemize}
\medskip

The computation of the one-loop corrections that we report here is not performed
by evaluating the integrals over two-point functions corresponding to
one-loop Feynman diagrams given below in Figs.~\ref{vac pol diagrams} 
and~\ref{2pt fun}.~\footnote{\linespread{1}\selectfont 
This would in fact not be possible as the photon propagators for power-law
inflation have not been worked out thus far.}
Instead, we formulate the problem in terms of double-differential 
equations descending solely from 
forming correlators of operator Maxwell's equations with themselves,
reminiscent of Dyson-Schwinger equations in the Schwinger-Keldysh 
formalism~\cite{Berges:2004yj,NeqQFT}.
Solving the equations directly in the superhorizon limit is straightforward,
and yields the one-loop correction to the superhorizon correlators.
A similar approach has already been employed 
in~\cite{Giovannini:2000dj,Calzetta:1997ku,Kandus:1999st}
when computing magnetic field correlators in the radiation era
 (see also~\cite{Kaya:2018qbj} for recent work on the topic).
This procedure greatly simplifies the technical difficulties 
of loop computations in inflation, if at the end of the day one is interested 
only in superhorizon behaviour, and potentially makes the 
two-loop computation feasible.

We find that all the one-loop E\&M correlators are  enhanced compared to
the tree-level vacuum fluctuations. The enhancement is manifested in three
aspects: (i) secular growth, (ii) spatial running, and (iii) IR sensitivity.
We find that in power-law inflation the secular growth of E\&M correlators is more pronounced than in de Sitter space, thus making the perturbation theory even less well behaved.
However, it is unclear exactly whether, or which aspect of the result ought to be considered 
as an artefact of time-dependent perturbation theory. Superhorizon
correlators also grow with comoving spatial separation, providing yet another 
limitation of perturbation theory. Finally, The IR sensitivity is manifested by the
almost logarithmic dependence of the correlators on the deep IR scale 
of the initial state, necessary to define the physical state of the complex scalar.
All three enhancements signal potentially large non-perturbative effects
for the E\&M correlators of massless SQED in power-law inflation, and
further investigation is clearly called for. 

Even though the magnitude of non-perturbative
effects is beyond the scope of this work, the one-loop computation reported here does offer an important insight.
The correlators that we compute, both for the
conserved~$U(1)$ current and for the E\&M fields,
satisfy distinct hierarchies. 
For current correlators the hierarchy between components is~$(ij) \!\gg\! (0i) \!\gg\! (00)$,
while the E\&M correlators the hierarchy is~$(EE) \!\gg\! (EB) \!\gg\! (BB)$.
Both of the hierarchies are in fact double hierarchies, satisfied by 
both the secular growth, and by the spatial running.
These hierarchies 
depend only on the homogeneous operator equations: the conservation equation for
the current and Faraday's law of induction for the E\&M fields, and they have to hold 
generally in inflating space-times even at the non-perturbative level.
In a decelerating universe after inflation Faraday's law~\cite{Kobayashi:2019uqs}
work towards inverting this hierarchy, leading to transient effects reported 
in~\cite{Kobayashi:2019uqs}.
The amplitude of the correlators, on the other hand, is not universal and depends on the 
specificities of the model and the background space-time. The linear fluctuations in 
the symmetry-breaking Abelian Higgs model in power-law
inflation considered recently in~\cite{Glavan:2020zne} 
demonstrates explicitly the hierarchy reported here.

\medskip

The remainder of the paper is organized as follows:
The following section introduces the power-law inflation space-time, and the 
SQED model in it. Sec.~\ref{sec: Quantum fluctuations} discusses our approach 
to quantifying quantum fluctuations -- computing correlators of charge currents
and EM fields, while Sec.~\ref{sec: Current correlators} and~\ref{sec: Field strength correlators},
are devoted to computing the said correlators. The general hierarchies between
the superhorizon correlators are discussed in Sec.~\ref{sec: Superhorizon hierarchies},
while the last section is reserved for the discussion of the results.

\section{Scalar electrodynamics in FLRW}
\label{sec: Scalar electrodynamics in FLRW}

In this section we briefly introduce the power-law inflation
space-time and the scalar quantum electrodynamics model living in it.

\subsection{FLRW and power-law inflation}
\label{subsec: FLRW and power-law inflation}

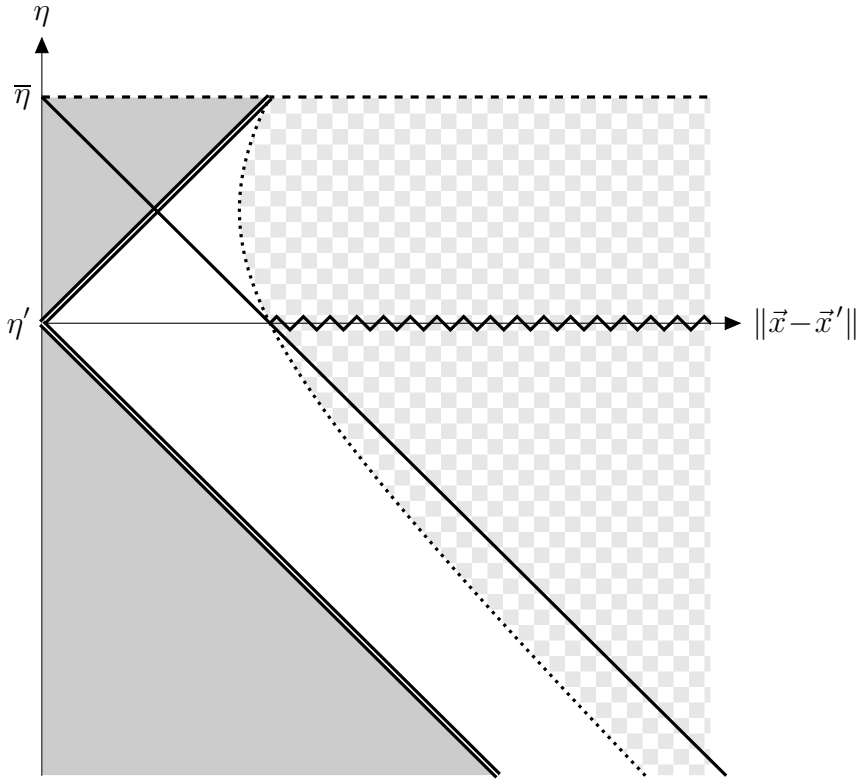
\begin{figure}[h!]
\centering
\begin{tikzpicture}

\draw[draw=white, fill=black!20] (0,-3) -- (3,0) -- (0,0) -- (0,-3);

\draw[draw=white, fill=black!20] (0,-3) -- (6,-9) -- (0,-9) -- (0,-3);

\draw[pattern=checkerboard, pattern color=black!10, draw=white,
	domain=-9:0,smooth,samples=100,variable=\y]  
	plot ({  sqrt(\y*\y+3*\y+9)  },{\y}) -- (8.8,0) -- (8.8,-9) -- cycle;

\draw[-triangle 45] (0,-9) -- (0,.8) ;
\node[align=left, above] at (0,0.8) {$\eta$} ;

\draw[-triangle 45] (0,-3) -- (9.2,-3) ;
\node[align=left, right] at (9.2,-3) {$\| \vec{x} \!-\! \vec{x}^{\,\prime} \|$} ;

\draw[very thick,dashed] (0,0) -- (8.8,0) ;
\node[align=left, left] at (0,0) {$\overline{\eta}$} ;

\node[align=left, left] at (0,-3) {$\eta'$} ;

\draw[very thick, double] (0,-3) -- (3,0) ;

\draw[very thick, double] (0,-3) -- (6,-9) ;

\draw[domain=-9:0,smooth,samples=100,variable=\y,dotted,very thick]  plot ({  sqrt(\y*\y+3*\y+9)  },{\y});

\draw[very thick,snake=zigzag] (3,-3) -- (8.8,-3);

\draw[very thick] (0,0) -- (9,-9) ;

\end{tikzpicture}
%
\caption{\linespread{1}\selectfont 
Conformal diagram of power-law inflation space-time.
Two double lines represent the past and the future null rays of some 
point~$(\eta',\vec{x}^{\,\prime})$. The gray shaded region represents its past
and future light-cones,
and from the allowed range of
coordinates in Eq.~(\ref{coordinate range}), it follows that the past light-cone can 
grow without limits, while the future
light cone is limited by the asymptotic future time~$\overline{\eta}$. 
The dotted curve along with the checkered region represent generalized
superhorizon separations ($\overline{\rm SH}$) from Eq.~(\ref{generalized SH}),
while the jagged (zig-zag) line represents the spatial points $\left(\eta',\vec{x}\right)$ at superhorizon (SH) separations
from~$(\eta',\vec{x}^{\,\prime})$.
The bold solid line denotes the Hubble horizon of the comoving observer at~$\vec{x}^{\,\prime}$.
}
\label{conformal_diagram}
\end{figure}

The invariant line element of spatially flat Friedmann-Lema\^{i}tre-Robertson-Walker (FLRW)
space-time in conformal coordinates is given by,
\begin{equation}
ds^2 = a^2(\eta) \bigl[ - d\eta^2 + d\vec{x}^{\, 2} \bigr] \, ,
\end{equation}
where~$a(\eta)$ is the scale factor expressed in terms of conformal time~$\eta$,
and hence the metric is conformally flat,~$g_{\mu\nu} \!=\! a^2 \eta_{\mu\nu}$,
where the Minkowski metric is~$\eta_{\mu\nu} \!=\! \text{diag}(-1,1,1,1)$.
A quantity which encodes the rate of change of the scale factor
is the {\it conformal Hubble rate},
\begin{equation}
\mathcal{H} = \frac{\partial_0 a}{a} \, ,
\end{equation}
which is related to the physical Hubble rate~$H$ as~$\mathcal{H}\!=\!aH$.
Acceleration of the expansion is conveniently encoded by the so-called
{\it principal slow-roll parameter},
\begin{equation}
\epsilon \equiv 1 - \frac{\partial_0 \mathcal{H}}{\mathcal{H}^2} \, ,
\end{equation}
related to the often used deceleration parameter~$q$ as~$\epsilon\!=\!q\!+\!1$.
When~$\epsilon\!<\!1$ the expansion is accelerating, and when~$\epsilon\!>\!1$
it is decelerating.
In primordial inflation it typically takes small values $0\!<\! \epsilon \! \ll \! 1$. 
In this work we consider the special class
of FLRW space-times -- {\it power-law inflation} -- characterized by the
constant principal slow-roll parameter,
\begin{equation}
\epsilon = \text{const.}
\qquad \qquad
0 \le \epsilon < 1
\end{equation}
The scale factor and the conformal Hubble rate in power-law inflation are
related as~$\mathcal{H} \!=\! H_0 a^{1-\epsilon}$, and 
have particular time dependencies,
\begin{equation}
a(\eta) = \Bigl[ 1 - (1\!-\!\epsilon) H_0 (\eta\!-\!\eta_0) \Bigr]^{\frac{-1}{1-\epsilon}} \, ,
\qquad \qquad
\mathcal{H}(\eta) = H_0  \Bigl[ 1 - (1\!-\!\epsilon) H_0 (\eta\!-\!\eta_0) \Bigr]^{-1} \, ,
\label{power-law inflation quantities}
\end{equation}
where~$a(\eta_0)\!=\!1$ and~$\mathcal{H}(\eta_0) \!=\! H_0$. 
The ranges of coordinates in power-law inflation are,
\begin{equation}
- \infty < \eta < \overline{\eta} \equiv \eta_0 + \frac{1}{(1\!-\!\epsilon)H_0} \, ,
\qquad \qquad
- \infty < x^i < \infty \, ,
\label{coordinate range}
\end{equation}
and the conformal diagram of the space-time is given in Fig.~\ref{conformal_diagram}.
In this paper we shall consider a more restricted class of power-law inflation
for which~$0 \!<\! \epsilon \! < \! 1/2$ for the sake of simplicity, which is also
appropriate for inflation applications where~$0 \!<\! \epsilon \! \ll \! 1$.

\

In this work we mostly deal with correlators -- bi-local objects dependent on two points
of the manifold which we denote by~$x$ and~$x'$. Henceforth,
in bi-local objects, primed quantities are assumed dependent on the primed coordinate~$x'$,
and while unprimed quantities are assumed dependent on the unprimed
coordinate~$x$, {\it} e.g.~$\mathcal{H}' \!=\! \mathcal{H}(\eta')$,
and~$\mathcal{H}\!=\! \mathcal{H}(\eta)$.
It is convenient for us to define the standard notion of superhorizon separation between two
points~$x$ and~$x'$ at equal time slices,
\begin{equation}
\text{SH: }
\qquad \qquad
\| \Delta \vec{x} \| \gg \frac{1}{(1\!-\!\epsilon)\mathcal{H}} \, ,
\qquad \qquad
\eta'=\eta \, ,
\label{SH def}
\end{equation}
and also the notion of generalized superhorizon separation for non-equal times,
\begin{equation}
\overline{\text{SH}}: 
\qquad \qquad
\| \Delta \vec{x} \|^2 - (\eta \!-\! \eta')^2 
	\gg \frac{1}{(1\!-\!\epsilon)^2 \mathcal{H} \mathcal{H}'} \, ,
\label{generalized SH}
\end{equation}
where~$\Delta\vec{x}\!=\! \vec{x} \!-\! \vec{x}^{\,\prime}$.
The latter is introduced to allow us to take time derivatives of bi-local functions
in intermediate steps of the computations, but at the end we are interested in
the time-coincidence limit.
The symbols~$\stackrel{\rm \scr SH}{\sim}$ 
and~$\stackrel{\rm \scr \overline{SH}}{\sim}$
will denote the two limits, respectively.

\subsection{SQED in FLRW}
\label{subsec: SQED in FLRW}

The action for scalar electrodynamics in a 4-dimensional curved 
space-time is given by,
\begin{equation}
S[A_\mu, \Phi, \Phi^*] =
	\Lint d^{4\!}x \, \sqrt{-g} \, \biggl[
	- \frac{1}{4} g^{\mu\rho} g^{\nu\sigma} F_{\mu\nu} F_{\rho\sigma}
	- g^{\mu\nu} \bigl( D_\mu^* \Phi^* \bigr) \bigl( D_\nu \Phi \bigr)
	\biggr] \, ,
\end{equation}
where~$F_{\mu\nu} \!=\! \partial_\mu A_\nu \!-\! \partial_\nu A_\mu$
is the vector field strength tensor,~$D_{\mu} \!=\! \nabla_\mu \!+\! i q A_\mu$
is the~$U(1)$ covariant derivative,~$q$ is the coupling constant (the $U(1)$ charge),
and~$g\!=\!{\rm det}(g_{\mu\nu})$
is the metric determinant.
The equations of motion for the classical theory descending from this action are,
\begin{equation}
g^{\rho\sigma} \nabla_\rho F_{\sigma \mu}
	 =q  J_\mu \, ,
\qquad \qquad
g^{\mu\nu} D_\mu D_\nu \Phi = 0 \, ,
\label{EOMs}
\end{equation}
where the source for the Maxwell's equation above is the~$U(1)$ 
Noether's current,
\begin{equation}
J_\mu 
	= i \Bigl[ \bigl( D_\mu^* \Phi^* \bigr) \Phi - \Phi^*  \bigl( D_\mu \Phi \bigr)  \Bigr]
	= i \Bigl[ \bigl( \partial_\mu \Phi^* \bigr) \Phi - \Phi^*  \bigl( \partial_\mu \Phi \bigr)  \Bigr]
		+ 2 q A_\mu \Phi^* \Phi \, ,
\label{J intro}
\end{equation}
which is covariantly conserved,
\begin{equation}
g^{\mu\nu} \nabla_\mu J_\nu = 0 \, ,
\end{equation}
as a consequence of the second (scalar field) equation. Note that~$\Phi^*$
satisfies the complex conjugate of the second equation in~(\ref{EOMs}).
In addition to dynamical equations in~(\ref{EOMs}), the vector field strength
also satisfies homogeneous Maxwell's equations simply on the account of
its anti-symmetry,~$\nabla_{[\mu} F_{\nu\rho]}\!=\!0$.

\medskip

The dynamics of the quantum theory (scalar quantum electrodynamics -- SQED) 
is captured by the equations of motion
formally of the same form as the classical equations~(\ref{EOMs}), where the 
fields are substituted by the field operators,
\begin{equation}
A_\mu \rightarrow \hat{A}_\mu \, ,
\qquad \qquad
\Phi \rightarrow \hat{\Phi} \, ,
\qquad \qquad
\Phi^* \rightarrow \hat{\Phi}^\dag \, ,
\label{field operators}
\end{equation}
everywhere in the equations, including the covariant derivatives,
and the conserved current in~(\ref{J intro}). In addition,
the field operators in~(\ref{field operators}) satisfy canonical commutation relations,
which for gauge theories such as SQED depend on the choice of gauge.
Even though our results are gauge-invariant, for definiteness we will be
working in the exact covariant gauge,~\footnote{\linespread{1}\selectfont 
This gauge condition should be thought of as the~$\xi\!\rightarrow\!0$
limit of the general covariant averaged gauge, given by the gauge-fixing action,
\begin{equation*}
S_{\rm gf}[A_\mu] = \int \! d^{4\!}x \, \sqrt{-g} \, \biggl[
	- \frac{1}{2\xi} \bigl( g^{\mu\nu} \nabla_\mu A_\nu \bigr)^{\!2}
	\biggr] \, ,
\end{equation*}
which is appropriate for the indefinite metric quantization.
}
\begin{equation}
g^{\mu\nu} \nabla_\mu \hat{A}_\nu \!=\! 0 \, ,
\end{equation}
which is implicit in the operator version of equations of motion~(\ref{EOMs}).

\section{Quantum fluctuations}
\label{sec: Quantum fluctuations}

We are interested in computing the properties of electric and magnetic fields in 
a globally (and locally) neutral state, that contains no condensates of electric
and magnetic fields,
\begin{equation}
\bigl\langle \hat{J}_\mu(x) \bigr\rangle = 0 \, ,
\qquad \qquad
\bigl\langle \hat{F}_{\mu\nu}(x) \bigr\rangle = 0 \, .
\end{equation}
Even though the state has no net charge, 
because of the fundamental quantum nature of the fields they do exhibit
quantum fluctuations, {\it i.e.} correlators of 
currents and field strengths do not vanish,
\begin{equation}
\bigl\langle \hat{J}_\mu(x) \, \hat{J}_\nu(x') \bigr\rangle \neq 0 \, ,
\qquad \qquad
\bigl\langle \hat{F}_{\mu\nu}(x) \, \hat{F}_{\rho\sigma}(x') \bigr\rangle \neq 0 \, .
\label{fluctuations}
\end{equation}
These correlators are what we wish to compute in time coincidence and for 
superhorizon distances, as defined in~(\ref{SH def}). 
In addition to being physically relevant, the superhorizon limit allows us to 
simplify the computation considerably.
Instead of going through the standard diagrammatic approach 
to one-loop computations, we choose
to present the computation in a more intuitive framework that ties
better to way things are thought of in cosmology.

\medskip

The tree-level correlator of vector field strength, in absence of coupling to
charged fields in power-law inflation (and in general FLRW) is given by
\begin{equation}
\bigl\langle \hat{F}_{\mu\nu}(x) \, \hat{F}_{\rho\sigma}(x') \bigr\rangle
	= \frac{2}{\pi^2 \bigl( \Delta x^2 \bigr)^2} \biggl[
		\eta_{\mu [\rho} \eta_{\sigma] \nu} 
	- 4 \eta_{\alpha[\mu} \eta_{\nu] [\sigma} \eta_{\rho]\beta} 
		\frac{ \Delta x^\alpha \Delta x^\beta}{\Delta x^2} \biggr] \, ,
\label{F tree correlator}
\end{equation}
where~$\Delta x^\mu \!=\! x^\mu \!-\! x'^\mu$, 
and~$\Delta x^2 \!=\! \eta_{\mu\nu} \Delta x^\mu \Delta x^\nu$. This correlator 
takes the same form as in flat space due to: (i) vector field being conformally
coupled to gravity in~$D\!=\!4$, (ii) FLRW space-time being conformally flat,
and (iii) defining a conformally invariant state for the vector field.

\medskip

The vector field in SQED is sourced by the complex scalar via the~$U(1)$ current.
The operator versions of Maxwell equations take the form,
\begin{equation}
g^{\alpha\beta} \nabla_\alpha \hat{F}_{\beta \mu}
	 = q \hat{J}_\mu \, ,
\qquad \qquad
\nabla_{[\alpha} \hat{F}_{\mu\nu]} = 0 \, ,
\label{covariant Maxwell}
\end{equation}
where the first one is an operator version of the first equation in~(\ref{EOMs}), 
while the second one is a consequence of the anti-symmetry 
of the fields strength tensor,
and where bracketed indices denote weighted anti-symmetrization.
We shall combine these operator equations in a way that allows us 
to compute the vector field strength correlators conveniently.
Taking the correlator of the first of these equations with itself results in an equation
relating the correlators in~(\ref{fluctuations}), while taking the correlator of the
second equation with the field strength tensor results in the homogeneous equation
for the second correlator in~(\ref{fluctuations}),
\begin{align}
&
\bigl[ g^{\alpha\rho}(x) \nabla_\alpha \bigr] \times 
	\bigl[ g^{\beta\sigma}(x') \nabla'_\beta \bigr] \times
	\bigl\langle \hat{F}_{\rho\mu}(x) \, \hat{F}_{\sigma\nu}(x') \bigr\rangle
	= q^2 \bigl\langle \hat{J}_\mu(x) \, \hat{J}_\nu(x') \bigr\rangle \, ,
\label{inhomogeneous}
\\
&
\nabla_{[\alpha} \bigl\langle \hat{F}_{\mu\nu]}(x) \, \hat{F}_{\rho\sigma}(x') \bigr\rangle = 0 \, ,
\qquad \quad\quad
\nabla'_{\beta]} \bigl\langle \hat{F}_{\mu\nu}(x) \, \hat{F}_{[\rho\sigma}(x') \bigr\rangle = 0 \, .
\label{homogeneous}
\end{align}
These equations are of a type similar to Dyson-Schwinger equations
for Wightman functions.
They are perfectly suited for our purposes of perturbative computation:
the source on the right hand side of~(\ref{inhomogeneous})
-- the~$U(1)$ charge current correlator --
comes multiplied by the expansion parameter~$q^2$. Moreover,
they allow for an immediate physical interpretation: the tree-level 
correlator of charge currents sources the one-loop correction to the field strength 
correlator. This is precisely the way in which we are going to perform the computation: 
we shall compute the tree-level current correlator induced on the scalar by the inflating
space-time in Sec.~\ref{sec: Current correlators}, 
and then we shall use that result to correct the electric and magnetic field 
correlators in Sec.~\ref{sec: Field strength correlators}.


\section{Current correlators}
\label{sec: Current correlators}

The operator of the conserved~$U(1)$ current is given by~(\ref{J intro}),
where the fields are substituted by field operators,
\begin{equation}
\hat{J}_\mu
	= i \Bigl[ \hat{\Phi} \bigl( \partial_\mu \hat{\Phi}^\dag \bigr) 
		- \hat{\Phi}^\dag \bigl( \partial_\mu \hat{\Phi} \bigr)
		 \Bigr]
	+ 2 q \hat{A}_\mu \, \hat{\Phi}^\dag \, \hat{\Phi} \, ,
\label{J operator}
\end{equation}
and it is covariantly conserved as an operator identity,
\begin{equation}
0 = g^{\mu\nu} \nabla_\mu \hat{J}_\nu \, .
\label{current operator conservation}
\end{equation}
Here we wish to compute the correlators of the components of the~$U(1)$
current, $\bigl\langle \hat{J}_\mu(x) \, \hat{J}_\nu(x') \bigr\rangle$ with 
respect to the locally (and globally) neutral state, {\it i.e.} the state
in which the expectation value of the current 
vanishes,~$\bigl\langle \hat{J}_\mu (x) \bigr\rangle = 0$.
Even though neutral on average, that state still experiences quantum fluctuations, 
and hence the local
charge also must fluctuate. This effect is captured by the components
of the current correlators.
We compute these to leading order in the coupling constant,
which corresponds to neglecting its coupling to the
vector. For the globally neutral state
the leading order current correlators reduce to
\begin{equation}
\bigl\langle \hat{J}_\mu(x) \, \hat{J}_\nu(x') \bigr\rangle
	= 2 \, i \Delta(x;x') \times \partial_\mu \partial_\nu' i \Delta(x;x')
	- 2 \, \partial_\mu i \Delta(x;x') \times \partial_\nu' i \Delta(x;x') \, ,
\label{general correlator}
\end{equation}
where~$i\Delta(x;x')$ is the 2-point function of the free complex scalar 
field,~\footnote{\linespread{1}\selectfont 
The 2-point function we consider has no time ordering in its definition,
{\it i.e.} it is a Wightman rather than the Feynman 2-point function. However,
as we are interested in correlators on superhorizon distances, using either 
suffices, and we do not distinguish between them explicitly.}
\begin{equation}
i \Delta(x;x') = \bigl\langle \hat{\Phi}^\dag(x) \, \hat{\Phi}(x') \bigr\rangle \, .
\label{2pt function}
\end{equation}
For a discussion on charge fluctuations in primordial inflation and their possible
effects during and after inflation
see~\cite{Giovannini:2000dj,Calzetta:1997ku,DOnofrio:2012qeh,Goolsby-Cole:2015chd}.

When the scalar two-point function (propagator) in~(\ref{2pt function}) is known 
then it is a simple matter of acting on it with derivatives to obtain the 
charge current correlators in~(\ref{general correlator}). However, the propagator
for the minimally coupled massless scalar
is not known for arbitrary FLRW space-times.~\footnote{\linespread{1}\selectfont 
In the case of the conformally
coupled scalar the propagator is known for all the FLRW space-times
since it is not sensitive to the expansion, and takes the 
form of the rescaled flat space propagator.} One notable example is the power-law 
inflation space-time introduced in Sec.~\ref{subsec: FLRW and power-law inflation},
for which scalar propagator has been reported in~\cite{Janssen:2008px}.
There the propagator for the Chernikov-Tagirov-Bunch-Davies (CTBD) 
state~\cite{Chernikov:1968zm,Bunch:1978yq},
defined as the state that minimizes energy mode-per-mode in the asymptotic past, 
was computed in~$D$ space-time dimensions. The well known IR divergence of the
exact CTBD state in accelerating space-times~\cite{Ford:1977in,Janssen:2009nz}
(see also~\cite{Lochan:2018pzs} for a recent review)
is regulated by assuming that the IR modes below some scale~$k_0\!\ll\!H_0$ 
have a suppressed spectrum compared to the CTBD one, and only the leading 
contribution in~$k_0$ is kept.
Here we need the~$D\!=\!4$ limit of the propagator from~\cite{Janssen:2008px} which reads,
\begin{align}
i \Delta(x;x')
	={}&
	\biggl[  \frac{(1\!-\!\epsilon) H_0 }{4\pi} \biggr]^{2}
	\Biggl\{
	\Gamma\bigl( \tfrac{3}{2} \!+\! \nu \bigr) 
			\, \Gamma\bigl( \tfrac{3}{2} \!-\! \nu \bigr)
			 (aa')^{-\epsilon} \times
	{}_2 F_1 \Bigl( \bigl\{ \tfrac{3}{2} \!+\! \nu , \tfrac{3}{2} \!-\! \nu \bigr\} ,
		\bigl\{ 2 \bigr\} , 1 \!-\! \tfrac{y}{4} \Bigr)
\nonumber \\
&	\hspace{4cm}
	+ \frac{\Gamma^2(2\nu) }
		{(2\nu \!-\! 3) \, \Gamma^2\bigl( \frac{1}{2} \!+\! \nu \bigr) }
	\biggl[ \frac{(1\!-\!\epsilon)H_0}{2 k_0} \biggr]^{2\nu - 3 }
	\Biggr\} \, ,
\label{scalar propagator}
\end{align}
%
where we have introduced the {\it distance function},
\begin{equation}
y(x;x') = (1\!-\!\epsilon)^2 \mathcal{H} \mathcal{H}'
	\Bigl[ \| \Delta\vec{x} \|^2 
		- \bigl(  \eta\!-\! \eta' \bigr)^{\!2} \, \Bigr] \, ,
\label{dS distance}
\end{equation}
with~$\Delta\vec{x} \!=\! \vec{x} \!-\! \vec{x}^{\,\prime} $, the parameters and their ranges are,
\begin{equation}
\nu = \frac{(3\!-\!\epsilon)}{2(1\!-\!\epsilon)} \, ,
\qquad \qquad
\frac{3}{2} \le \nu < \frac{5}{2} 
\quad \text{for} \quad 0\le \epsilon < \frac{1}{2} \, .
\end{equation}
and where~$k_0$
is the effective IR regulator,~\footnote{\linespread{1}\selectfont 
The scalar propagator in~(\ref{scalar propagator}) is valid 
for~$k_0\| \Delta\vec{x} \| \!\ll\!1$.
Beyond this region its form changes and it typically decays exponentially with
spatial distance, depending on the specifics of the deep IR sector of 
the state.} satisfying~$0 \!<\! k_0 \!\ll\! (1\!-\!\epsilon) H_0$.
The dependence on the scale~$k_0$ physically implies vast IR gravitational particle production,
and~$k_0$ should properly be seen as dependence on the initial conditions.

As we are interested in computing the correlators at superhorizon distances
we will only need the asymptotic form of the propagator,
\begin{align}
\MoveEqLeft[3]
i \Delta(x;x') 
	\stackrel{\rm \scr \overline{SH}}{\sim}
	\biggl[  \frac{(1\!-\!\epsilon) H_0 }{4\pi} \biggr]^{\!2}
	\frac{2^{3-2\nu} \, \Gamma(2\nu)}{\Gamma\bigl( \tfrac{1}{2} \!+\! \nu \bigr)}
	\times \Biggl\{
	\frac{\Gamma(2\nu)}{(2\nu \!-\! 3) \, \Gamma\bigl( \tfrac{1}{2} \!+\! \nu \bigr)}
	\biggl[  \frac{(1\!-\!\epsilon) H_0}{k_0} \biggr]^{2\nu-3}
\nonumber \\
&	\hspace{0.cm}
	+ \bigl[  (1\!-\!\epsilon) H_0 \| \Delta\vec{x} \| \Bigr]^{2\nu-3}
	\Biggl[ \Gamma\bigl( \tfrac{3}{2} \!-\! \nu \bigr)
	+ \Gamma\bigl( \tfrac{5}{2} \!-\! \nu \bigr)
	\Bigl( \frac{\mathcal{H}}{\mathcal{H}'} 
		\!+\! \frac{\mathcal{H}'}{\mathcal{H}} \Bigr)
		\frac{1}{(1\!-\!\epsilon)^2 \mathcal{H} \mathcal{H}' \| \Delta\vec{x} \|^2}
\nonumber \\
&	\hspace{0.cm}
	+ \Gamma\bigl( \tfrac{7}{2} \!-\! \nu \bigr)
	\biggl( \frac{1}{2} \Bigl( \frac{\mathcal{H}}{\mathcal{H}'}
		\!+\! \frac{\mathcal{H}'}{\mathcal{H}} \Bigr)^{\!2}
		- \frac{1}{\nu\!-\!1} \biggr)
	\frac{1}{(1\!-\!\epsilon)^4 (\mathcal{H} \mathcal{H}')^2 \| \Delta\vec{x} \|^4} \Biggr]
	\Biggr\} \, .
\end{align}

\medskip

Plugging this expansion into~(\ref{general correlator}) yields the desired 
leading order contributions to superhorizon current correlators,
\begin{align}
\bigl\langle \hat{J}_0(x) \, \hat{J}_0(x') \bigr\rangle
	\stackrel{\rm \scr \overline{SH}}{\sim} {}& 
	\mathcal{N} \biggl[  \frac{(1\!-\!\epsilon) H_0 }{4\pi} \biggr]^{\!4}
	\times
	\frac{\bigl[ (1\!-\!\epsilon)H_0 \| \Delta\vec{x} \| \bigr]^{4\nu-6}}
		{\| \Delta\vec{x} \|^2}
	\times
	\frac{1}{(1\!-\!\epsilon)^2 \mathcal{H} \mathcal{H}' \| \Delta\vec{x} \|^2 }
\label{J0 J0}
\\
&	\hspace{1.2cm} 
		\times \frac{(\nu\!-\!2)  }{(\nu\!-\!1) }
	\Biggl[
	\frac{1}{\bigl( k_0 \| \Delta\vec{x} \| \bigr)^{2\nu-3} }
	+ \frac{2(4\nu \!-\! 7) \, \Gamma\bigl( \tfrac{5}{2}\!-\!\nu \bigr) \, 
				\Gamma\bigl( \tfrac{1}{2} \!+\! \nu \bigr)}
		{(2\nu\!-\!5)(\nu\!-\!2) \, \Gamma(2\nu)}
	\Biggr] \, ,
\nonumber 
\\
\bigl\langle \hat{J}_0(x) \, \hat{J}_i(x') \bigr\rangle
	\stackrel{\rm \scr \overline{SH}}{\sim} {}& 
	- \mathcal{N}  \biggl[  \frac{(1\!-\!\epsilon) H_0 }{4\pi} \biggr]^{\!4}
	\times
	\frac{\bigl[ (1\!-\!\epsilon) H_0 \| \Delta \vec{x} \| \bigr]^{4\nu-6}}
		{ \| \Delta\vec{x} \|^2}
	\times
	\frac{1}{(1\!-\!\epsilon) \mathcal{H} \| \Delta\vec{x} \|  }
\label{J0 Ji}
\\
&
	\hspace{1.2cm}
	\times \Biggl[ 
		\frac{1}{\bigl( k_0 \| \Delta\vec{x} \| \bigr)^{2\nu-3} }
		+ \frac{4 \, \Gamma\bigl( \tfrac{5}{2} \!-\! \nu \bigr) \, 
				\Gamma\bigl( \tfrac{1}{2} \!+\! \nu \bigr)}
			{(2\nu \!-\! 5) \, \Gamma(2\nu)} 
		\Biggr] \times \frac{\Delta x_i}{\| \Delta \vec{x} \|} \, ,
\nonumber
\\
\bigl\langle \hat{J}_i(x) \, \hat{J}_j(x') \bigr\rangle
	\stackrel{\rm \scr \overline{SH}}{\sim} {}& 
	- \mathcal{N} \biggl[  \frac{(1\!-\!\epsilon) H_0 }{4\pi} \biggr]^{\!4}
	\times
	\frac{\bigl[ (1\!-\!\epsilon) H_0 \| \Delta\vec{x} \| \bigr]^{4\nu-6}} {\| \Delta \vec{x} \|^2 } 
	\times
\label{Ji Ji}
\\
&
\hspace{1.2cm}
\times
\Biggl\{ 
	\frac{1}{(2\nu \!-\! 5)} \Biggl[ 
		\frac{1}{\bigl( k_0 \| \Delta\vec{x} \| \bigr)^{2\nu-3}}
		- \frac{2 \,\Gamma\bigl( \tfrac{5}{2} \!-\! \nu \bigr) \,\Gamma\bigl( \tfrac{1}{2} \!+\! \nu \bigr)}
				{\Gamma(2\nu)}
		\Biggr] \delta_{ij}
\nonumber \\
&
\hspace{2.cm}
	+ \Biggl[ \frac{1}{\bigl( k_0 \| \Delta\vec{x} \| \bigr)^{2\nu-3}}
		+ \frac{4 \,\Gamma\bigl( \tfrac{5}{2} \!-\! \nu \bigr) \,\Gamma\bigl( \tfrac{1}{2} \!+\! \nu \bigr)}
				{(2\nu\!-\!5) \, \Gamma(2\nu)}
		\Biggr] 
		\frac{\Delta x_i \, \Delta x_j}{\| \Delta \vec{x} \|^2} 
	\Biggr\} \, , 
\nonumber
\end{align}
where the overall normalization constant is
\begin{equation}
\mathcal{N} = 
	\frac{8 \, \Gamma(2\nu) \, \Gamma\bigl( \tfrac{7}{2} \!-\! \nu \bigr)}
		{ (2\nu\!-\!3) \, \Gamma\bigl( \tfrac{1}{2} \!+\! \nu \bigr) }
		\biggl[ \frac{2^{3-2\nu} \, \Gamma(2\nu)}{\Gamma\bigl( \tfrac{1}{2} \!+\! \nu \bigr)} \biggr]^2 \, .
\end{equation}
The conservation of the current operator~(\ref{current operator conservation})
implies consistency conditions on both legs of the current correlators,
\begin{align}
0 ={}&
	- \partial_0 \bigl\langle \hat{J}_0(x) \, \hat{J}_\mu(x') \bigr\rangle
	- 2 \mathcal{H} \bigl\langle \hat{J}_0(x) \, \hat{J}_\mu(x') \bigr\rangle
	+ \partial_i \bigl\langle \hat{J}_i(x) \, \hat{J}_\mu(x') \bigr\rangle \, ,
\label{conservation 1}
\\
0 ={}&
	- \partial_0' \bigl\langle \hat{J}_\mu(x) \, \hat{J}_0(x') \bigr\rangle
	- 2 \mathcal{H}' \bigl\langle \hat{J}_\mu(x) \, \hat{J}_0(x') \bigr\rangle
	+ \partial_i' \bigl\langle \hat{J}_\mu(x) \, \hat{J}_i(x') \bigr\rangle \, ,
\label{conservation 2}
\end{align}
which are exactly satisfied by~(\ref{J0 J0})--(\ref{Ji Ji}).
Note that, apart from the numerical coefficients and the tensor structures,
at late times, and large spatial separations, the $(00)$ correlator is down
by a factor~$\bigl( \mathcal{H}' \| \Delta\vec{x} \| \bigr)$ compared to the~$(0i)$
correlator, which is 
down by a factor~$\bigl( \mathcal{H}\| \Delta\vec{x} \| \bigr)$
compared to the~$(ij)$ correlator. This hierarchy is a generic feature 
expounded on in Sec.~\ref{sec: Superhorizon hierarchies}.

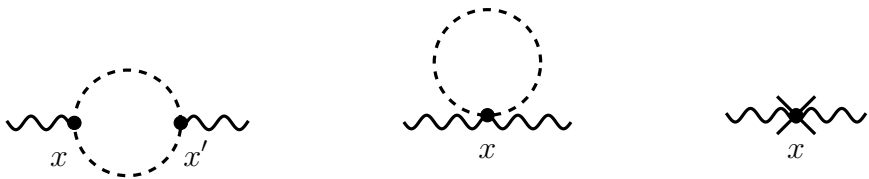
\begin{figure}[h!]
\vspace{4mm}
\begin{equation*}
\begin{tikzpicture}[baseline={0cm-0.5*height("$=$")}]
\draw[very thick,dashed] (0,0) circle (0.7) ;
\filldraw (-0.7,0) circle (2.5pt) node {} ;
\filldraw (0.7,0) circle (2.5pt) node {} ;

\draw[very thick] plot[domain=0.7:1.52, samples=300] (\x+0.07,{0.09*sin( 5 * pi * (\x - 0.7) r )});
\draw[very thick] plot[domain=-0.7:-1.52, samples=300] (\x-0.07,{-0.09*sin( 5 * pi * (\x + 0.7) r )});

\node at (-0.9,-0.475) {$x$};
\node at (0.9,-0.4) {$x'$};

\end{tikzpicture}
\hspace{2cm}
\begin{tikzpicture}[baseline={-0.1cm-0.5*height("$=$")}]
\draw[very thick,dashed] (0,0.7) circle (0.7) ;
\filldraw (0,0) circle (2.5pt) node {} ;

\draw[very thick] plot[domain=-1.1:1.1, samples=300] (\x,{-0.09+0.09*cos( 5 * pi * \x r )});

\node at (0,-0.475) {$x$};
\end{tikzpicture}
\hspace{2cm}
\begin{tikzpicture}[baseline={-0.1cm-0.5*height("$=$")}]
\filldraw (0,0) circle (2.5pt) node {} ;

\draw[very thick] plot[domain=-0.92:0.92, samples=300] (\x,{-0.09*cos( 5 * pi * \x r )});

\node at (0,-0.475) {$x$};

\draw[very thick] (-0.25,-0.25) -- (0.25,0.25) ;
\draw[very thick] (-0.25,0.25) -- (0.25,-0.25) ;
\end{tikzpicture}
\end{equation*}
\vskip-0.3cm
\caption{\linespread{1}\selectfont 
Diagrams contributing to full one-loop vacuum polarization: {\it left:} bubble diagram,
{\it middle:} seagull diagram, {\it right:} photon field strength 
renormalization counterterm. 
Dashed lines represent the scalar propagator, and wavy lines the
amputated vector propagator. The endpoints are assumed to carry 
a space-time label, a vector index, and a Schwinger-Keldysh polarity label.
}
\label{vac pol diagrams}
\end{figure}

In the language of the Schwinger-Keldysh formalism  of QFT,
the charge correlators we have computed correspond to
the $(-+)$ component of the one-loop vacuum polarization,
\begin{equation}
i \bigl[ \tensor*[_\mu^{\scr - }]{\Pi}{_\nu^{\scr +} } \bigr](x;x') 
	= q^2 (aa')^2 \bigl\langle \hat{J}_\mu(x) \, \hat{J}_\nu(x') \bigr\rangle \, ,
\label{vac pol definition}
\end{equation}
which at superhorizon distances also corresponds to the $(++)$ component,
\begin{equation}
i \bigl[ \tensor*[_\mu^{\scr + }]{\Pi}{_\nu^{\scr +} } \bigr](x;x') 
	=
	 q^2 (aa')^2 \bigl\langle \mathcal{T} \bigl[  \hat{J}_\mu(x) \, \hat{J}_\nu(x') \bigr] \bigr\rangle 
	 \stackrel{\rm \scr \overline{SH}}{\sim}
	 q^2 (aa')^2 \bigl\langle \hat{J}_\mu(x) \, \hat{J}_\nu(x') \bigr\rangle \, ,
\end{equation}
where~$\mathcal{T}$ denotes time-ordering.
The factors of the scale factor are due to the definition of vacuum polarization.
The diagrams corresponding to the one-loop vacuum polarization are given
in Fig.~\ref{vac pol diagrams}. 
It is only the non-local diagram that contributes to the~$(-+)$ component,
while all three contribute to the~$(++)$ component. However, the two
local diagrams contribute only in coincidence,~$\propto\!\delta^4(x\!-\!x')$,
and not for superhorizon separations. We neglect their contributions
on two accounts: (i) it was shown in~\cite{Prokopec:2002uw,Prokopec:2003tm} that in de Sitter space 
the only local contribution
contributing to the vacuum polarization is the conformal anomaly,
whose effect is weak~\cite{Benevides:2018mwx}, and (ii) we expect them to contribute
only subdominantly as they correspond to homogeneous solutions of 
Eqs.~(\ref{important eq. 3})--(\ref{important eq. 3}) discussed in the following section.

Determining whether correlators~(\ref{J0 J0})--(\ref{Ji Ji}) represent a big or a small 
effect is not as straightforward since (i) there is no lower-order effect to compare to,
and (ii) the redshifting of the correlators depends on whether we present them with
raised or lowered indices. A more meaningful comparison can be done on the level of electric 
and magnetic field correlators given in Sec.~\ref{sec: Field strength correlators}, where one can compare them to tree-level contribution~(\ref{F tree correlator}). 

\medskip


The exact de Sitter limit,~$\epsilon\!\to\!0$, of the current correlators~(\ref{Ji Ji})--(\ref{J0 J0})
takes a simpler form,
\begin{align}
\bigl\langle \hat{J}_0(x) \, \hat{J}_0(x') \bigr\rangle
	\stackrel{\rm \scr \overline{SH}}{\sim} {}& 
	\frac{H_0^4}{8\pi^4 \| \Delta\vec{x} \|^2}
	 \times
		\frac{1}{\mathcal{H} \mathcal{H}' \| \Delta\vec{x} \|^2}
	\times
	\Bigl[ - 3 + 2 \gamma_{\scr \rm E} + 2 \ln \bigl( k_0 \| \Delta\vec{x} \| \bigr) \Bigr] \, ,
\label{dS J0 J0}
\\
\bigl\langle \hat{J}_0(x) \, \hat{J}_i(x') \bigr\rangle
	\stackrel{\rm \scr \overline{SH}}{\sim} {}& 
	\frac{H_0^4}{8\pi^4 \| \Delta\vec{x} \|^2}
	 \times
		\frac{1}{\mathcal{H} \| \Delta\vec{x} \|}
		\times
	\Bigl[ - 2 \ln \bigl( k_0 \| \Delta\vec{x} \| \bigr)- 2 \gamma_{\scr \rm E} + 1 \, \Bigr] 
		\frac{\Delta x_i}{\| \Delta \vec{x} \|} \, ,
\label{dS J0 Ji}
\\
\bigl\langle \hat{J}_i(x) \, \hat{J}_j(x') \bigr\rangle
	\stackrel{\rm \scr \overline{SH}}{\sim} {}& 
	\frac{H_0^4}{8\pi^4 \| \Delta\vec{x} \|^2 }
	\Biggl\{ 
	 \Bigl[ - \ln\bigl( k_0 \| \Delta\vec{x} \| \bigr) - \gamma_{\scr \rm E} + 1 \, \Bigr]  \delta_{ij}
 \nonumber \\
&
\hspace{3.0cm}
	+ \Bigl[ 2 \ln \bigl( k_0 \| \Delta\vec{x} \| \bigr) + 2 \gamma_{\scr \rm E} - 1 \, \Bigr] 
		\frac{\Delta x_i \, \Delta x_j}{\| \Delta \vec{x} \|^2} 
	\Biggr\} \, ,
\label{dS Ji Ji}
\end{align}
where~$\gamma_{\scr \rm E}$ is the Euler-Mascheroni constant.
It provides an important consistency check with the previously reported results,
as it reproduces the superhorizon limit of the vacuum polarization
from~\cite{Prokopec:2002uw} (Eq.~(73) from their paper), provided that we make 
the identification~$k_0\!=\! H_0 \, e^{1-\gamma_{\scr \rm E}}$, and taking the 
factors from~(\ref{vac pol definition}) into account.

\section{Field strength correlators}
\label{sec: Field strength correlators}

Charge current correlators are not the best object to characterize the physical
behaviour. At tree-level they are zero so we have nothing to compare with
when asking how big they are. More appropriate quantities are correlators of 
vector field strength tensor,
\begin{equation}
\bigl \langle \hat{F}_{\mu\nu}(x) \, \hat{F}_{\rho \sigma}(x') \bigr\rangle \, ,
\end{equation}
where~$\hat{F}_{\mu\nu} \!=\! \partial_\mu \hat{A}_\nu \!-\! \partial_\nu \hat{A}_\mu$.
It is better to present this correlator in the form of electric and magnetic
field correlators, where the E\&M field operators are, respectively,
\begin{equation}
\hat{E}_i = \hat{F}_{0i} \, ,
\qquad \qquad
\hat{B}_i = - \frac{1}{2} \varepsilon_{ijk} \hat{F}_{jk} \, ,
\label{EB def}
\end{equation}
and where~$\varepsilon_{ijk}$ is the 3-dimensional Levi-Civita symbol.
The E\&M correlators exist even in the free theory because of the quantum fluctuations.
Due to the conformal coupling,
the free vector field does not sense the expansion of the FLRW space-time,
and the electric and magnetic field correlators take the same form~(\ref{F tree correlator}) as
in flat space,
\begin{align}
&
\bigl\langle \hat{E}_i(\eta,\vec{x}) \, \hat{E}_j(\eta,\vec{x}^{\,\prime}) \bigr\rangle
	= \frac{1}{\pi^2 \| \Delta\vec{x} \|^4} \biggl[
		- \delta_{i j} + 2 \frac{\Delta x_{i} \Delta x_{j} }{\| \Delta \vec{x} \|^2} \biggr] \, ,
\\
&
\bigl\langle \hat{E}_i(\eta,\vec{x}) \, \hat{B}_j(\eta,\vec{x}^{\,\prime}) \bigr\rangle
	= 0
\\
&
\bigl\langle \hat{B}_i(\eta,\vec{x}) \, \hat{B}_j(\eta,\vec{x}^{\,\prime}) \bigr\rangle
	= \frac{1}{\pi^2 \| \Delta\vec{x} \|^4} \biggl[
		- \delta_{i j} + 2 \frac{\Delta x_{i} \Delta x_{j} }{\| \Delta \vec{x} \|^2} \biggr] \, .
\end{align}
As a consequence of coupling to the charged scalar this does not remain true --
the scalar mediates the effects of the expansion to the vector field.

We quantify the influence of the charged scalar on the vector
by computing the one-loop correction to E\&M field correlators,
which is the first correction in the coupling constant~$q^2$. 
This we do by solving perturbatively 
a variant of the equations~(\ref{inhomogeneous}) and~(\ref{homogeneous}) for the correlators 
derived in Sec.~\ref{sec: Quantum fluctuations}. Instead of the covariant version,
it is better to write them in terms of E\&M fields, for which the
Maxwell's equations read,
\begin{align}
&
\partial_i \hat{E}_i = - a^2 q \hat{J}_0 \, ,
&&
\partial_0 \hat{E}_i - \varepsilon_{ijk} \partial_j \hat{B}_k = - a^2 q \hat{J}_i   \, .
\label{inhomogeneous Maxwell}
\\
&
\partial_i \hat{B}_i = 0 \, ,
&&
\partial_0 \hat{B}_i + 
\varepsilon_{ijk} \partial_j \hat{E}_k
= 0  \, .
\label{homogeneous Maxwell}
\end{align}
The first line corresponds to the inhomogeneous equation in~(\ref{covariant Maxwell})
containing respectively Gauss' law and Amp\`{e}re's law,
while the second line corresponds to the homogeneous equations in~(\ref{covariant Maxwell})
containing respectively the law of vanishing magnetic charge, and
Faraday's law. Instead of forming the correlators of these first order equations 
directly, it is more convenient to use the second-order sourced wave equations,
\begin{align}
&
\bigl( \partial_0^2 - \nabla^2 \bigr) \hat{E}_i
	=
	- \partial_0 \bigl( a^2 q \hat{J}_i \bigr)
	+ a^2 \partial_i q \hat{J}_0 \, ,
\\
&
\bigl( \partial_0^2 - \nabla^2 \bigr) \hat{B}_i
	= \varepsilon_{ijk} \partial_j a^2 q \hat{J}_k \, ,
\end{align}
in which E\&M fields decouple. Forming the correlators between different combinations
of the two equations above results in,
\begingroup
\allowdisplaybreaks
\begin{align}
&
\bigl( \partial_0^2 - \nabla^2 \bigr) \bigl( \partial_0'^2 - \nabla'^2 \bigr)
	\bigl\langle \hat{E}_i(x) \, \hat{E}_j(x') \bigr\rangle
\nonumber \\
&	\hspace{1.8cm}
	= q^2 \partial_0 \partial_0' \Bigl[ (aa')^2 \bigl\langle \hat{J}_i(x) \, \hat{J}_j(x') \bigr\rangle \Bigr]
	- q^2 \partial_0 \partial_j' \Bigl[ (aa')^2 \bigl\langle \hat{J}_i(x) \, \hat{J}_0(x') \bigr\rangle \Bigr]
 \label{important eq. 1} \\
&	\hspace{3.3cm}
	- q^2 \partial_i \partial_0'  \Bigl[ (aa')^2 \bigl\langle \hat{J}_0(x) \, \hat{J}_j(x') \bigr\rangle \Bigr]
	+ q^2 \partial_i \partial_j'  \Bigl[ (aa')^2 \bigl\langle \hat{J}_0(x) \, \hat{J}_0(x') \bigr\rangle \Bigr]
 \, ,
\nonumber
\\
&
\bigl( \partial_0^2 - \nabla^2 \bigr) \bigl( \partial_0'^2 - \nabla'^2 \bigr)
	\bigl\langle \hat{E}_i(x) \, \hat{B}_j(x') \bigr\rangle
\label{important eq. 2} \\
&	\hspace{1.8cm}
	= - q^2 \varepsilon_{jkl} \, \partial_0 \partial_k'
		\Bigl[ (aa')^2 \bigl\langle \hat{J}_i(x) \, \hat{J}_l(x') \bigr\rangle \Bigr]
	+ q^2 \varepsilon_{jkl} \, \partial_i \partial_k'
		\Bigl[ (aa')^2 \bigl\langle \hat{J}_0(x) \, \hat{J}_l(x') \bigr\rangle \Bigr] \, ,
\nonumber
\\
&
\bigl( \partial_0^2 - \nabla^2 \bigr) \bigl( \partial_0'^2 - \nabla'^2 \bigr)
	\bigl\langle \hat{B}_i(x) \, \hat{B}_j(x') \bigr\rangle
\nonumber \\
&	\hspace{1.8cm}
	= q^2 \varepsilon_{ikl} \, \varepsilon_{jmn} \, \partial_k \partial'_m
	\Bigl[ (aa')^2 \bigl\langle \hat{J}_l(x) \, \hat{J}_n(x') \bigr\rangle \Bigr] 
 \, ,
 \label{important eq. 3} 
\end{align}
\endgroup
The utility of these particular equations becomes apparent when we apply them
to solving for superhorizon correlators. Firstly, note that in the superhorizon limit a spatial 
derivative introduces another factor of~$1/\| \Delta\vec{x} \|$, while a time derivative introduces 
another factor of~$\mathcal{H}$, and therefore the spatial derivatives on the right hand side
of equations above can be dropped in favour of time derivatives.
Secondly, all current
correlators on the right hand side, except for the~$(ij)$ component, may be dropped 
due to the superhorizon hierarchy between them discussed in 
Sec.~\ref{sec: Current correlators},
\begin{align}
&
\bigl( \partial_0 \partial_0' \bigr)^2 \bigl\langle \hat{E}_i(x) \, \hat{E}_j(x') \bigr\rangle
	\stackrel{\rm \scr \overline{SH}}{\sim}
	q^2 \partial_0 \partial_0' \Bigl[ (aa')^2
		\bigl\langle \hat{J}_i(x) \, \hat{J}_j(x') \bigr\rangle \Bigr] \, ,
\label{redEq1}
\\
&
\bigl( \partial_0 \partial_0' \bigr)^2 \bigl\langle \hat{E}_i(x) \, \hat{B}_j(x') \bigr\rangle
	\stackrel{\rm \scr \overline{SH}}{\sim}
	- q^2 \varepsilon_{jkl} \, \partial_0 \partial_k'
		\Bigl[ (aa')^2 \bigl\langle \hat{J}_i(x) \, \hat{J}_l(x') \bigr\rangle \Bigr] \, ,
\label{redEq2}
\\
&
\bigl( \partial_0 \partial_0' \bigr)^2 \bigl\langle \hat{B}_i(x) \, \hat{B}_j(x') \bigr\rangle
	\stackrel{\rm \scr \overline{SH}}{\sim}
	q^2 \varepsilon_{ikl} \, \varepsilon_{jmn}  \, \partial_k \partial'_m
	\Bigl[ (aa')^2 \bigl\langle \hat{J}_l(x) \, \hat{J}_n(x') \bigr\rangle \Bigr] \, .
\label{redEq3}
\end{align}
Furthermore, the $(ij)$ component~(\ref{Ji Ji}) of the current correlator
is time-independent, making the task of inverting the time derivatives on the 
left-hand side straightforward. Making use of~(\ref{power-law inflation quantities}),
\begin{align}
&
\bigl\langle \hat{E}_i(x) \, \hat{E}_j(x') \bigr\rangle
	\stackrel{\rm \scr \overline{SH}}{\sim}
	q^2 \frac{ (aa')^{1+\epsilon} }{(1\!+\!\epsilon)^2 H_0^2} \,
		\bigl\langle \hat{J}_i(x) \, \hat{J}_j(x') \bigr\rangle \, ,
\label{EE corr}
\\
&
\bigl\langle \hat{E}_i(x) \, \hat{B}_j(x') \bigr\rangle
	\stackrel{\rm \scr \overline{SH}}{\sim}
	- q^2 \frac{ a^{1+\epsilon} (a')^{2\epsilon}  }{2\epsilon(1\!+\!\epsilon)^2 H_0^3} \,
		\varepsilon_{jkl} \, \partial_k'
		\bigl\langle \hat{J}_i(x) \, \hat{J}_l(x') \bigr\rangle \, ,
\label{EB corr}
\\
&
\bigl\langle \hat{B}_i(x) \, \hat{B}_j(x') \bigr\rangle
	\stackrel{\rm \scr \overline{SH}}{\sim}
	q^2 \frac{(aa')^{2\epsilon} }
			{4 \epsilon^2 (1\!+\!\epsilon)^2 H_0^4} \,
		\varepsilon_{ikl} \varepsilon_{jmn} \, \partial_k \partial_m'
		\bigl\langle \hat{J}_l(x) \, \hat{J}_n(x') \bigr\rangle \, .
\label{BB corr}
\end{align}
Note that these expressions are valid for late times in power-law inflation,
when~$\ln(a)\!>\!1/\epsilon$. The de Sitter limit~$\epsilon\!\to\!0$ cannot be taken directly
in these late-time results on the account of factors of~$\epsilon$ in the denominator.
This tells us that the subleading term in the expansion, which is negligible
at late times in power-law inflation, becomes equally relevant in this limit.
However, instead of having to work out the subleading terms, it is simpler to 
invert equations~(\ref{redEq1})--(\ref{redEq3}) in the de Sitter limit directly, which we do
at the end of this section.



The solutions~(\ref{EE corr})--(\ref{BB corr}) pass the consistency checks as they satisfy
the various relations descending from the linear Maxwell 
equations~(\ref{inhomogeneous Maxwell}) and~(\ref{homogeneous Maxwell}).
In particular relations descending from the inhomogeneous equations,
\begin{align}
&
q^2 (aa')^2 \bigl\langle \hat{J}_0(x) \, \hat{J}_0(x') \bigr\rangle
	\stackrel{\rm \scr \overline{SH}}{\sim}
		 \partial_i \partial_j'  
		\bigl\langle \hat{E}_i(x) \, \hat{E}_j(x') \bigr\rangle  \, ,
\\
&
q^2 (aa')^2 \bigl\langle \hat{J}_0(x) \, \hat{J}_i(x') \bigr\rangle 
	\stackrel{\rm \scr \overline{SH}}{\sim}
	\partial_j \partial_0' \bigl\langle \hat{E}_j(x) \, \hat{E}_i(x') \bigr\rangle
	+ \partial_j \epsilon_{ikl} \partial_k'
		\bigl\langle \hat{E}_j(x) \, \hat{B}_l(x') \bigr\rangle \, ,
\\
&
q^2 (aa')^2 \bigl\langle \hat{J}_i(x) \, \hat{J}_j(x') \bigr\rangle 
	\stackrel{\rm \scr \overline{SH}}{\sim}
	\partial_0 \partial_0' \bigl\langle \hat{E}_i(x) \, \hat{E}_j(x') \bigr\rangle
	+ \epsilon_{jmn} \partial_0 \partial_m' 
		\bigl\langle \hat{E}_i(x) \, \hat{B}_n(x') \bigr\rangle
 \\
&	\hspace{4.2cm}
	+ \epsilon_{ikl} \partial_k \partial_0' 
		\bigl\langle \hat{B}_l(x) \, \hat{E}_j(x') \bigr\rangle
	+ \epsilon_{ikl} \epsilon_{jmn} \partial_k \partial_m'
		\bigl\langle \hat{B}_l(x) \, \hat{B}_n(x') \bigr\rangle \, ,
\nonumber
\end{align}
are satisfied to leading order,
and the relations descending from homogeneous equations,
\begin{align}
0 ={}&
\varepsilon_{jkl} \partial_k' \bigl\langle \hat{E}_i(x) \, \hat{E}_l(x') \bigr\rangle
	+ \partial_0' \bigl\langle \hat{E}_i(x) \, \hat{B}_j(x') \rangle \, ,
\\
0 ={}&
\varepsilon_{ikl} \partial_k \bigl\langle \hat{E}_l(x) \, \hat{B}_j(x') \bigr\rangle
	+ \partial_0 \bigl\langle \hat{B}_i(x) \, \hat{B}_j(x') \rangle \, ,
\\
0 ={}&
\partial_j' \bigl\langle \hat{E}_i(x) \, \hat{B}_j(x') \bigr\rangle \, ,
\\
0 ={}&
\partial_i \bigl\langle \hat{B}_i(x) \, \hat{B}_j(x') \bigr\rangle \, ,
\end{align}
are satisfied exactly.
Additional relations can be generated by exploiting symmetries of the correlators
under the excange of indices and ordering of the fields.

\medskip

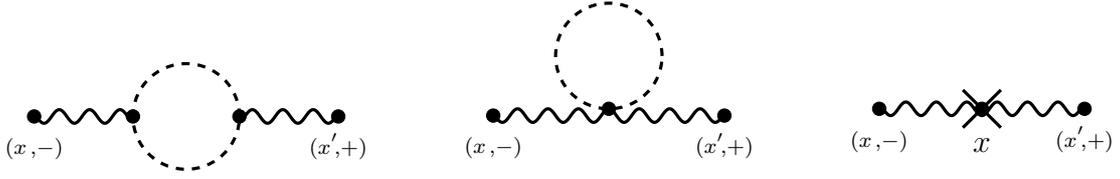
\begin{figure}[h!]
\begin{equation*}
\begin{tikzpicture}[baseline={0cm-0.5*height("$=$")}]
\draw[very thick,dashed] (0,0) circle (0.7) ;
\filldraw (-0.7,0) circle (2.5pt) node {} ;
\filldraw (0.7,0) circle (2.5pt) node {} ;
\filldraw (-2.0,0) circle (2.5pt) node {} ;
\filldraw (2.0,0) circle (2.5pt) node {} ;

\draw[very thick] plot[domain=0.7:1.95, samples=300] (\x+0.07,{0.09*sin( 5 * pi * (\x - 0.7) r )});
\draw[very thick] plot[domain=-0.7:-1.95, samples=300] (\x-0.07,{-0.09*sin( 5 * pi * (\x + 0.7) r )});

\node at (-2.,-0.4) {$\scriptstyle(x^{\color{white} \prime} \!, -)$};
\node at (2.,-0.4) {$\scriptstyle(x'\!, +)$};

\end{tikzpicture}
\hspace{1.cm}
\begin{tikzpicture}[baseline={-0.1cm-0.5*height("$=$")}]
\draw[very thick,dashed] (0,0.7) circle (0.7) ;
\filldraw (0,0) circle (2.5pt) node {} ;

\draw[very thick] plot[domain=-1.52:1.52, samples=300] (\x,{-0.09+0.09*cos( 5 * pi * \x r )});

\filldraw (1.52,-0.09) circle (2.5pt) node {} ;
\filldraw (-1.52,-0.09) circle (2.5pt) node {} ;

\node at (-1.52,-0.49) {$\scriptstyle(x^{\color{white} \prime} \!, -)$};
\node at (1.52,-0.49) {$\scriptstyle(x'\!, +)$};
\end{tikzpicture}
\hspace{1.cm}
\begin{tikzpicture}[baseline={-0.1cm-0.5*height("$=$")}]
\filldraw (0,0) circle (2.5pt) node {} ;

\draw[very thick] plot[domain=-1.35:1.35, samples=300] (\x,{-0.09*cos( 5 * pi * \x r )});

\filldraw (1.35,0) circle (2.5pt) node {} ;
\filldraw (-1.35,0) circle (2.5pt) node {} ;

\node at (0,-0.475) {$x$};

\node at (-1.35,-0.4) {$\scriptstyle(x^{\color{white} \prime} \!, -)$};
\node at (1.35,-0.4) {$\scriptstyle(x'\!, +)$};

\draw[very thick] (-0.25,-0.25) -- (0.25,0.25) ;
\draw[very thick] (-0.25,0.25) -- (0.25,-0.25) ;
\end{tikzpicture}
\end{equation*}
\vskip-0.3cm
\caption{\linespread{1}\selectfont 
Diagrams contributing to one-loop corrections 
of the vector field two-point function~$\bigl\langle \hat{A}_\mu(x) \, \hat{A}_\nu(x') \bigr\rangle$;
{\it left:} bubble diagram,
{\it middle:} seagull diagram, 
{\it right:} photon field strength 
renormalization counterterm.
Dashed lines represent the tree-level scalar propagator, and wavy lines the
photon propagator.
The Schwinger-Keldysh polarities of the endpoints denote these are the diagams 
corresponding to the Wightman two-point function.}
\label{2pt fun}
\end{figure}

The computation presented here is equivalent to the standard QFT computation,
where we would first compute the one-loop correction to the Wightman
two-point function of the vector field~$\bigl\langle \hat{A}_\mu(x) \hat{A}_\nu(x') \bigr\rangle$,
given by the diagrams in Fig.~\ref{2pt fun},
and then act with derivative operators to obtain the E\&M correlators,
\begin{equation}
\bigl\langle \hat{F}_{\mu\nu}(x) \, \hat{F}_{\rho\sigma}(x') \bigr\rangle
	= 4 \bigl( \partial_{[\mu} \delta^\alpha_{\nu]} \bigr)
		\bigl( \partial'_{[\rho} \delta^\beta_{\sigma]} \bigr)
	\bigl\langle \hat{A}_\alpha(x) \hat{A}_\beta(x') \bigr\rangle \, .
\end{equation}
As we are only interested in superhorizon behaviour, this way we have avoided
having to go through a more complicated computation of first computing a
gauge-dependent object and only then acting with derivative operators
to project out the physical information. In fact, it is not possible to perform
the dimensionally regulated one-loop computation in power-law inflation, 
since the photon propagators have not been reported in $D$-dimensional 
power-law inflation, and would have to be computed first.
Our computation captures the effects of the first diagram in Fig.~\ref{2pt fun},
which is the important one at superhorizon separations, and represents the reaction of
the photon to the charge current fluctuations of the complex scalar induced by the 
expansion of the space-time (see Fig.~\ref{charged part prod}). 
The explicit expressions for the one-loop corrected equal-time correlators 
of electric and magnetic fields at super-Hubble separations follow from~(\ref{EE corr})--(\ref{BB corr}),
\begin{align}
&
\bigl\langle \hat{E}_i(\eta,\vec{x}) \, \hat{E}_j(\eta,\vec{x}^{\,\prime}) \bigr\rangle
		\stackrel{\rm \scr SH}{\sim}
\nonumber \\
&	\hspace{1.2cm}
	\frac{1}{\pi^2 \| \Delta\vec{x} \|^4} \Biggl\{
	\delta_{ij} \biggl[ 
		-1 - q^2
		\bigl[ (1\!-\!\epsilon) \mathcal{H} \| \Delta\vec{x} \| \bigr]^{4\nu-4}
		\mathcal{K}_1 \bigl( k_0 \| \Delta\vec{x} \| \bigr)
		 \biggr] 
	\, ,
\label{EE final}
\\
&	\hspace{4cm}
+ \frac{\Delta x_i \Delta x_j}{\| \Delta\vec{x} \|^2} \biggl[
	2 - q^2 \bigl[ (1\!-\!\epsilon) \mathcal{H} \| \Delta\vec{x} \| \bigr]^{4\nu-4}
		\mathcal{K}_2 \bigl( k_0 \| \Delta\vec{x} \| \bigr)
	\biggr]
	\Biggr\} \, ,
\nonumber 
\\
&
\bigl\langle \hat{E}_i(\eta,\vec{x}) \, \hat{B}_j(\eta,\vec{x}^{\,\prime}) \bigr\rangle
		\stackrel{\rm \scr SH}{\sim}
\nonumber \\
&	\hspace{1.2cm}
	\frac{1}{\pi^2 \| \Delta\vec{x} \|^4} \Biggl\{
	\varepsilon_{ijk} \frac{\Delta x_k}{\| \Delta\vec{x} \|}
		\biggl[ 0 - 
			\frac{ q^2 }{2\epsilon}
			\bigl[ (1\!-\!\epsilon) \mathcal{H} \| \Delta\vec{x} \| \bigr]^{4\nu-5} \biggr]
			\mathcal{K}_3\bigl( k_0 \| \Delta\vec{x} \| \bigr)
	\Biggr\}
	\, ,
\label{EB final}
\\
&
\bigl\langle \hat{B}_i(\eta,\vec{x}) \, \hat{B}_j(\eta,\vec{x}^{\,\prime}) \bigr\rangle
		\stackrel{\rm \scr SH}{\sim}
\nonumber \\
&	\hspace{1.2cm}	
	\frac{1}{\pi^2 \| \Delta\vec{x} \|^4} \Biggl\{
	\delta_{ij} \biggl[ -1 + \frac{ q^2 }{4\epsilon^2}
		\bigl[ (1\!-\!\epsilon) \mathcal{H} \| \Delta\vec{x} \| \bigr]^{4\nu-6}
		\mathcal{K}_4\bigl( k_0 \| \Delta\vec{x} \| \bigr)  \biggr]
\label{BB final} \\
&	\hspace{4cm}
	+ \frac{\Delta x_i \Delta x_j}{\| \Delta\vec{x} \|^2} \biggl[
	2 + \frac{ q^2 }{4\epsilon^2}
		\bigl[ (1\!-\!\epsilon) \mathcal{H} \| \Delta\vec{x} \| \bigr]^{4\nu-6}
		\mathcal{K}_5\bigl( k_0 \| \Delta\vec{x} \| \bigr)
	\biggr]
	\Biggr\}
	 \, ,
\nonumber
\end{align}
where the short hand notation is,
\begingroup
\allowdisplaybreaks
\begin{align}
\mathcal{K}_1 \bigl(k_0 \| \Delta\vec{x} \| \bigr) 
	={}& \frac{\mathcal{N}}{2^8 \pi^2 } 
		\Bigl( \frac{1\!-\!\epsilon}{1\!+\!\epsilon} \Bigr)^{\!2} 
		\frac{1}{(2\nu\!-\!5)} \Biggl[ \frac{1}{\bigl( k_0 \| \Delta\vec{x} \| \bigr)^{2\nu-3}}
		- \frac{2 \, \Gamma\bigl( \frac{5}{2} \!-\! \nu \bigr) \,
			\Gamma\bigl( \frac{1}{2} \!+\! \nu \bigr)}{\Gamma(2\nu)} \Biggr] \, ,
\\
\mathcal{K}_2 \bigl(k_0 \| \Delta\vec{x} \| \bigr) 
	={}& \frac{\mathcal{N}}{2^8 \pi^2 } 
		\Bigl( \frac{1\!-\!\epsilon}{1\!+\!\epsilon} \Bigr)^{\!2} 
		\biggl[
		 \frac{1}{\bigl( k_0 \| \Delta\vec{x} \| \bigr)^{2\nu-3}}
		+ \frac{4 \, \Gamma\bigl( \frac{5}{2} \!-\! \nu \bigr) \,
			\Gamma\bigl( \frac{1}{2} \!+\! \nu \bigr)}{(2\nu \!-\! 5) \, \Gamma(2\nu)}
	\biggr] \, ,
\\
\mathcal{K}_3 \bigl( k_0 \| \Delta\vec{x} \| \bigr)
	={}&	
		 \frac{(1\!-\!\epsilon)\mathcal{N}}{2^8 \pi^2 } 
		\Bigl( \frac{1\!-\!\epsilon}{1\!+\!\epsilon} \Bigr)^{\!2} 
		\biggl[
		 (4\nu\!-\!8) \mathcal{K}_1 + k_0 \| \Delta\vec{x} \| \mathcal{K}_1'
				- \mathcal{K}_2 
				\biggr] \, ,
\\
\mathcal{K}_4 \bigl( k_0 \| \Delta\vec{x} \| \bigr)
	={}&
		 \frac{(1\!-\!\epsilon)\mathcal{N}}{2^8 \pi^2 } 
		\Bigl( \frac{1\!-\!\epsilon}{1\!+\!\epsilon} \Bigr)^{\!2} 
		\biggl[
		(4\nu\!-\!8) \mathcal{K}_3 + k_0 \| \Delta\vec{x} \| \mathcal{K}'_3
		\biggr]	\, ,
 \\
\mathcal{K}_5 \bigl( k_0 \| \Delta\vec{x} \| \bigr)
	={}&	
		 \frac{(1\!-\!\epsilon)\mathcal{N}}{2^8 \pi^2 } 
		\Bigl( \frac{1\!-\!\epsilon}{1\!+\!\epsilon} \Bigr)^{\!2} 
		\biggl[
		- (4\nu\!-\!10) \mathcal{K}_3 - k_0 \| \Delta\vec{x} \| \mathcal{K}'_3
		\biggr]	\, ,
\end{align}
\endgroup
and where primes on~$\mathcal{K}$'s here denote a derivative with respect to the argument.
Importantly, note that no matter how small the coupling constant~$q$ is, given enough
time the one-loop correction will dominate over the tree-level contribution due to
power-law secular growth. Moreover, note that one-loop corrections satisfy a hierarchy. Apart from
the numerical coefficients, and the tensor structures, the~$(BB)$ correlator is down by
a factor~$\bigl( \mathcal{H} \| \Delta\vec{x} \| \bigr)$ compared to the~$(EB)$
correlator, which is down by a factor~$\bigl( \mathcal{H} \| \Delta\vec{x} \| \bigr)$
compared to the~$(EE)$ correlator. We will see in Sec.~\ref{sec: Superhorizon hierarchies}
this is not a peculiarity of the model, but rather a generic feature of E\&M correlators
in inflation.

\medskip

We conclude this section by working out the correlators in the de Sitter limit~$\epsilon\!=\!0$.
Note that in power-law inflation, apart from the overall factor~$1/\| \Delta\vec{x} \|^4$
and the IR sensitivity~$k_0\| \Delta\vec{x} \|$, the corrections~(\ref{EE final})--(\ref{BB final}) 
depend on coordinates only as~$(\mathcal{H}\| \Delta\vec{x} \|)$. This must be the case 
in de Sitter space as well, where this quantity is proportional to the physical distance.
Assuming this dependence and going through the exercise of 
inverting Eqs.~(\ref{redEq1})--(\ref{redEq3}) for~$\epsilon\!=\!0$
produces,
\begin{align}
&
\bigl\langle \hat{E}_i(\eta,\vec{x}) \, \hat{E}_j(\eta,\vec{x}^{\,\prime}) \bigr\rangle
	\stackrel{\rm \scr SH}{\sim}
	\frac{1}{\pi^2 \| \Delta\vec{x} \|^4} \Biggl\{
	1 + \frac{q^2}{8\pi^2} \bigl( \mathcal{H} \| \Delta\vec{x} \| \bigr)^2
	\biggl[
		\Bigl( - \ln\bigl(k_0 \| \Delta\vec{x} \| \bigr) \!-\! \gamma_{\scr \rm E} \!+\! 1 \Bigr) \delta_{ij}
\nonumber \\
&	\hspace{3cm}
	+ \Bigl( 2 \ln\bigl(k_0 \| \Delta\vec{x} \| \bigr) \!+\! 2\gamma_{\scr \rm E} \!-\! 1 \Bigr)
	\frac{\Delta x_i \Delta x_j}{\| \Delta\vec{x} \|^2}
	\biggr]
	\Biggr\} \, ,
\label{EE final dS}
\\
&
\bigl\langle \hat{E}_i(\eta,\vec{x}) \, \hat{B}_j(\eta,\vec{x}^{\,\prime}) \bigr\rangle
	\stackrel{\rm \scr SH}{\sim}
	\frac{1}{\pi^2 \| \Delta\vec{x} \|^4} \Biggl\{ 0
	- \frac{q^2}{4\pi^2} \bigl( \mathcal{H} \| \Delta\vec{x} \| \bigr) 
		\ln\bigl(\mathcal{H} \| \Delta\vec{x} \| \bigr) 
		\times \varepsilon_{ijk} \frac{\Delta x_k}{\| \Delta\vec{x} \|}
	\Biggr\} \, ,
\label{EB final dS}
\\
&
\bigl\langle \hat{B}_i(\eta,\vec{x}) \, \hat{B}_j(\eta,\vec{x}^{\,\prime}) \bigr\rangle
	\stackrel{\rm \scr SH}{\sim}
	\frac{1}{\pi^2 \| \Delta\vec{x} \|^4} \Biggl\{ 1
	+ \frac{q^2 }{2\pi^2} \ln^2\bigl( \mathcal{H} \| \Delta\vec{x} \| \bigr) 
		\biggl[ \delta_{ij} - 2 \frac{\Delta x_i \Delta x_i}{ \| \Delta\vec{x} \|^2 } \biggr]
	\Biggr\} \, .
\label{BB final dS}
\end{align}
Note that the hierarchy here is supplemented by additional factors 
of~$\ln(\mathcal{H}\| \Delta\vec{x} \|)$. The above results in the de Sitter limit
can be checked independently by performing the standard one-loop computation 
for the vector field two-point function
in the Schwinger-Keldysh formalism, making use of the fully renormalized
one-loop self-energy computed in Ref.~\cite{Prokopec:2002uw}.
That computation is left for future work.

\section{Superhorizon hierarchies}
\label{sec: Superhorizon hierarchies}

The final results for the superhorizon correlators in both
Sec.~\ref{sec: Current correlators} and~\ref{sec: Field strength correlators}
exhibit two types of hierarchies when in time-coincidence~($\eta'\!=\!\eta$). 
Firstly, the secular late-time hierarchy,
where given enough time the hierarchy develops because of secular growth,
and secondly, the spatial running hierarchy, where the larger the spatial
separation, the larger the relative magnitude of correlators.
For the conserved current correlators the hierarchy (up to numerical factors
and tensor structures) 
takes the form,
\begin{align}
&
\bigl\langle \hat{J}_0(\eta,\vec{x}) \, \hat{J}_0(\eta,\vec{x}^{\,\prime}) \bigr\rangle
	\Big/
\bigl\langle \hat{J}_0(\eta,\vec{x}) \, \hat{J}_i(\eta,\vec{x}^{\,\prime}) \bigr\rangle
	\Big/
\bigl\langle \hat{J}_i(\eta,\vec{x}) \, \hat{J}_j(\eta,\vec{x}^{\,\prime}) \bigr\rangle
\label{J hierarchy}
\\
&	\hspace{8cm}
	\stackrel{\rm \scr SH }{\sim}
\
	1 
		\Big/
	\mathcal{H} \| \Delta\vec{x} \|
		\Big/
	\bigl( \mathcal{H} \| \Delta\vec{x} \| \bigr)^2 \, ,
\nonumber
\end{align}
and for the E\&M correlators,
\begin{align}
&
\bigl\langle \hat{E}_i(\eta,\vec{x}) \, \hat{E}_j(\eta,\vec{x}^{\,\prime}) \bigr\rangle
	\Big/
\bigl\langle \hat{E}_i(\eta,\vec{x}) \, \hat{B}_j(\eta,\vec{x}^{\,\prime}) \bigr\rangle
	\Big/
\bigl\langle \hat{B}_i(\eta,\vec{x}) \, \hat{B}_j(\eta,\vec{x}^{\,\prime}) \bigr\rangle
\label{EM hierarchy}
\\
&	\hspace{8cm}
	\stackrel{\rm \scr SH }{\sim}
\
	\bigl( \mathcal{H} \| \Delta\vec{x} \| \bigr)^2
		\Big/
	\mathcal{H} \| \Delta\vec{x} \|
		\Big/
	1 \, .
\nonumber
\end{align}
These hierarchies seem very similar, as the ratios are powers 
of~$\bigl(\mathcal{H} \| \Delta\vec{x} \| \bigr)$. However, they are actually independent
of each other and have different origins. The current hierarchies are a consequence of the
covariant conservation of the~$U(1)$ 
current~(\ref{current operator conservation}), 
while the E\&M hierarchies are a consequence
of Faraday's law~(\ref{homogeneous Maxwell}). 
In Sec.~\ref{sec: Current correlators} 
and~\ref{sec: Field strength correlators} these two laws given by
homogeneous equations
have been utilized as consistency checks of the solutions obtained from the
inhomogeneous equations. Here we demonstrate that it is these laws that are 
responsible for the hierarchies in the first place.

The relevance of Faraday's law 
for redshifting of superhorizon magnetic fields 
was pointed out recently in~\cite{Kobayashi:2019uqs}. There it
was found to lead to very different scaling of magnetic fields in 
the radiation-dominated epoch, 
provided there was a large electric field generated in inflation. 
While in decelerating periods of expansion Faraday's law tells us
something about the transients, in inflation it
is responsible for establishing a definite hierarchy between the E\&M correlators,
on the account of accelerated expansion of the background space-time.
The conservation equation does the same for current correlators.

The hierarchies are
not specific to the particular model we examined in this paper, but rather arise 
generically in inflation, and are expected to hold at the non-perturbative
level as well. The expressions we give are specific to power-law inflation,
but should get only small corrections in general slow-roll inflation.
On the other hand, the amplitude of the correlators
is not universal, and does depend on the model, etc.
In the following Sec.~\ref{subsec: Conserved currents}
and~\ref{subsec: EM fields}
we give a brief account on the general structure of hierarchies between
the current correlators, and between the E\&M correlators.
The detailed study is left for future work.

\subsection{Conserved currents}
\label{subsec: Conserved currents}

Provided that the covariant conservation of the currents is not broken by quantum effects,
we may use the conservation 
equation~(\ref{current operator conservation}) as an operator identity,
and we may form 2-point functions (correlators) of conserved current operators.
Let us assume that at superhorizon separations the $(00)$ current correlator has the 
following leading asymptotic behaviour,
\begin{equation}
\bigl\langle \hat{J}_0(x) \, \hat{J}_0(x') \bigr\rangle
	\stackrel{\rm \scr \overline{SH}}{\sim}
	\mathcal{N} \| \Delta\vec{x} \|^{\alpha}
		(aa')^{\beta} \, .
\end{equation}
It follows immediately solely from the conservation equation 
that the remaining current correlators are,
\begin{align}
&
\bigl\langle \hat{J}_0(x) \, \hat{J}_i(x') \bigr\rangle
	\stackrel{\rm \scr \overline{SH}}{\sim}
	\bigl\langle \hat{J}_0(x) \, \hat{J}_0(x') \bigr\rangle
	\times
	\biggl[ - \Bigl( \frac{2 \!+\! \beta}{3 \!+\! \alpha} \Bigr) 
		\bigl( \mathcal{H}' \| \Delta\vec{x} \| \bigr) \biggr]
		\times \frac{\Delta x_i}{\| \Delta\vec{x} \|}  \, ,
\\
&
\bigl\langle \hat{J}_i(x) \, \hat{J}_j(x') \bigr\rangle
	\stackrel{\rm \scr \overline{SH}}{\sim}
	\bigl\langle \hat{J}_0(x) \, \hat{J}_0(x') \bigr\rangle
	\times
	\biggl[ - \frac{(2\!+\!\beta)^2}{(2\!+\!\alpha)(3\!+\!\alpha)}
		\bigl( \mathcal{H}\mathcal{H}' \| \Delta\vec{x} \|^2 \bigr) \biggr]
\nonumber \\
&	\hspace{3cm}
		\times 
		\biggl[ \Bigl( 1 + \frac{(3\!+\!\alpha)(4\!+\!\alpha)}{(2\!+\!\beta)^2} \# \Bigr) \delta_{ij}
		- \frac{(2\!+\!\alpha)(3\!+\!\alpha)}{(2\!+\!\beta)^2} \# 
			\frac{\Delta x_i \Delta x_j}{\| \Delta\vec{x} \|^2} \biggr]  \, ,
\end{align}
where~$\#$ is an undetermined constant.
This is precisely the hierarchy observed in~(\ref{J0 J0})--(\ref{Ji Ji}) that we have derived
here without refering to the actual model, or making coupling constant expansions.

\subsection{E\&M fields}
\label{subsec: EM fields}

Let us assume that the magnetic field correlator at superhorizon 
separations behaves at leading order 
as,~\footnote{\linespread{1}\selectfont 
Note that this is not the most general scaling that can result in inflation,
since we had already found that in the de Sitter limit~(\ref{EE final dS})--(\ref{BB final dS})
the power-law behaviour turns into logarithmic one. However, generalizations
to include these cases are straightforward, and we do not consider such
details here.
}
\begin{equation}
\bigl\langle \hat{B}_i(x) \, \hat{B}_j(x') \bigr\rangle
	\stackrel{\rm \scr \overline{SH}}{\sim}
	\mathcal{N} \| \Delta\vec{x} \|^\alpha (aa')^\beta
	\times
	\biggl[ (\alpha\!+\!2) \delta_{ij} - \alpha \frac{\Delta x_i \Delta x_j}{\| \Delta\vec{x} \|^2} \biggr]
	\, ,
\end{equation}
such that it satisfies the condition of no magnetic charge from~(\ref{homogeneous Maxwell}).
Then the remaining E\&M correlators follow from the 
Faraday's law in~(\ref{homogeneous Maxwell}),
\begin{align}
&
\bigl\langle \hat{E}_i(x) \, \hat{B}_j(x') \bigr\rangle
	\stackrel{\rm \scr \overline{SH}}{\sim}
	\mathcal{N} (aa')^\beta \| \Delta\vec{x} \|^\alpha
	\times
	\Bigl[ - \beta \bigl( \mathcal{H} \| \Delta\vec{x} \| \bigr) \Bigr]
	\times
	\varepsilon_{ijk} \frac{\Delta x_k}{\| \Delta\vec{x} \|}
	\, ,
\\
&
\bigl\langle \hat{E}_i(x) \, \hat{E}_j(x') \bigr\rangle
	\stackrel{\rm \scr \overline{SH}}{\sim}
	\mathcal{N} (aa')^\beta \| \Delta\vec{x} \|^\alpha
	\times
	\Bigl[ - \beta \bigl( \mathcal{H} \mathcal{H}' \| \Delta\vec{x} \|^2 \bigr) \Bigr]
\\
&	\hspace{5cm}
	\times
		\biggl[ \# \delta_{ij}
		- \Bigl( \beta^2 - (2\!+\!\alpha) \# \Bigr) \frac{\Delta x_i \Delta x_j}{\| \Delta\vec{x} \|^2} \biggr] \, ,
\nonumber
\end{align}
where~$\#$ is an undetermined constant.
Thus, Faraday's law precisely leads to the superhorizon hierarchies observed 
in~(\ref{EE final})--(\ref{BB final}), which we have derived without assuming a
particular model, or any particular approximation scheme,
Therefore. this is a general feature of E\&M correlators at superhorizon separations
in inflation. 
Strictly speaking, the hierarchy in this section has been derived for power-law infation,
but we expect the results to hold with small corrections in general
power-law inflation. Recently the tree-level~E\&M of the Abelian Higgs model 
have been worked out in power-law inflation~\cite{Glavan:2020zne}, and they exactly
respect the super-Hubble hierarchy in~(\ref{EM hierarchy}).

\section{Discussion}
\label{sec: Discussion}

In this work we have examined how the vector in the massless
SQED model responds to the charged current fluctuations of the complex
scalar induced by the accelerating expansion of power-law inflation.
More precisely, the effects that we have taken into account are
\begin{itemize}
\item
The creation of $U(1)$ charge fluctuations of the complex scalar due to
the expansion,

\item
Reaction of the $U(1)$ vector field to the gravitationally created charge fluctuations of the complex scalar,
\end{itemize}
which constitute one-loop effects. The effect is quantified by the two-point functions of conserved 
currents~(\ref{J0 J0})--(\ref{Ji Ji}), which source  the two-point function of 
electric and magnetic field fluctuations, given
in~(\ref{EE final})--(\ref{BB final}); these constitute the main result of the paper.
The E\&M correlators exhibit three relevant properties,
\begin{itemize}
\item
Secular enhancement -- growth in time compared to the tree-level correlators

\item
Spatial running -- growth with spatial separation compared to the tree-level
correlators, 

\item
IR sensitivity -- dependence on the IR scale present in the scalar propagator.
\end{itemize}
All three point to potentially large non-perturbative IR effects.
One might wonder whether the secular enhancement and spatial running go hand-in-hand
with IR sensitivity, and would be removed by resumming the leading orders at each loop.
This, however, does not seem to be the case, since all the E\&M correlators
have the same dependence on the IR scale~$k_0$, but different secular and spatial running.

It is noteworthy that both the components of the
tree-level conserved current correlators~(\ref{J0 J0})--(\ref{Ji Ji}),
and the one-loop E\&M correlators~(\ref{EE final})--(\ref{BB final}) satisfy 
hierarchies,~(\ref{J hierarchy}) and~(\ref{EM hierarchy}), respectively.
Even though our explicit results pertain to the one-loop level, the hierarchies 
correlators satisfy must hold at the non-perturbative level as well, as they are 
consequences of operator versions of current conservation and Faraday's law.
They have recently been confirmed for the Abelian Higgs model in power-law 
inflation~\cite{Glavan:2020zne}.

The results presented here open up several directions for further investigation.
The large electric field fluctuations coherent over superhorizon separations 
might dynamically provide the electric field necessary for the occurrence of
the Schwinger mechanism -- production of charged pairs in external electric fields --
that has attracted attention 
recently~\cite{Kobayashi:2014zza,Frob:2014zka,Banyeres:2018aax}, more so due to possible 
imprints it might leave on the cosmological observables. Furthermore, technical 
simplifications brought about by casting the problem in terms of
double-differential equations~(\ref{important eq. 1})--(\ref{important eq. 3}) for correlators make two-loop computation far more feasible. These equations can be readily solved in the superhorizon limit, and eliminate the need of computing some of the integrals associated with Feynman diagrams.
At one-loop there was no need for regularization and renormalization 
since we examined off-coincident full two-point functions. At two-loop order we
likely require only renormalization of the one-loop vacuum polarization and scalar self-mass
in power-law inflation; both are eminently feasible computations that have thus far been performed only in de Sitter space~\cite{Prokopec:2002uw,Prokopec:2003tm,Kahya:2005kj}.

Finally, the properties of the correlators discussed in this work make a good case for the 
eventual breakdown of perturbation theory. 
The question then arises on whether this 
breakdown signifies the existence of large non-perturbative effects. This question can 
only be answered by going beyond perturbation theory in the IR, for example by setting up 
the full set of corresponding Starobinsky's stochastic equations, generalizing the approach 
of~\cite{Tsamis:2005hd, Moss:2016uix}, and examining both the vector and complex scalar. 
The Renormalization Group can then be used in this stochastic context to examine properties 
of the correlators at large 
spatial separations, see \cite{Prokopec:2017vxx} for 
an application to a test field in de Sitter. This will require adaptation of existing techniques.  
Furthermore, obtaining further perturbative results to higher loop orders, using the techniques 
discussed in this work, would be very useful as (i) they would establish how higher
order physical processes scale in time and with spatial separation, 
and (ii) would provide a consistency check that any resummation method has to reproduce. 
We hope to return to all these issues in future work.

\acknowledgments
D.~G.~would like to thank Takeshi Kobayashi for the
discussions on the subject during the visit to ICTP in Trieste.
This work was partially supported by the Fonds de la Recherche 
Scientifique -- FNRS under Grant IISN 4.4517.08 -- Theory of fundamental 
interactions, and by the STFC grant ST/P000371/1 -- Particles, Fields and Spacetime.

\bibliographystyle{JHEP}
\bibliography{EM_correlators_SQED_inflation.bib}

\providecommand{\href}[2]{#2}\begingroup\raggedright\begin{thebibliography}{10}

\bibitem{Arkani-Hamed:2015bza}
N.~Arkani-Hamed and J.~Maldacena, \emph{{Cosmological Collider Physics}},
  \href{https://arxiv.org/abs/1503.08043}{{\ttfamily 1503.08043}}.

\bibitem{Meerburg:2016zdz}
P.D.~Meerburg, M.~M\"unchmeyer, J.B.~Mu\~noz and X.~Chen, \emph{{Prospects for
  Cosmological Collider Physics}},
  \href{https://doi.org/10.1088/1475-7516/2017/03/050}{\emph{JCAP} {\bfseries
  03} (2017) 050} [\href{https://arxiv.org/abs/1610.06559}{{\ttfamily
  1610.06559}}].

\bibitem{Parker:1968mv}
L.~Parker, \emph{{Particle creation in expanding universes}},
  \href{https://doi.org/10.1103/PhysRevLett.21.562}{\emph{Phys. Rev. Lett.}
  {\bfseries 21} (1968) 562}.

\bibitem{Parker:1969au}
L.~Parker, \emph{{Quantized fields and particle creation in expanding
  universes. 1.}}, \href{https://doi.org/10.1103/PhysRev.183.1057}{\emph{Phys.
  Rev.} {\bfseries 183} (1969) 1057}.

\bibitem{Birrell:1982ix}
N.D.~Birrell and P.C.W.~Davies, \emph{{Quantum Fields in Curved Space}},
  Cambridge University Press, Cambridge, UK (1984),
  \href{https://doi.org/10.1017/CBO9780511622632}{10.1017/CBO9780511622632}.

\bibitem{Mukhanov:2007zz}
V.~Mukhanov and S.~Winitzki, \emph{{Introduction to Quantum Effects in
  Gravity}}, Cambridge University Press, Cambridge, UK (2007),
  \href{https://doi.org/10.1017/CBO9780511809149}{10.1017/CBO9780511809149}.

\bibitem{Parker:2009uva}
L.E.~Parker and D.~Toms, \emph{{Quantum Field Theory in Curved Spacetime}:
  {Quantized Field and Gravity}}, Cambridge University Press, Cambridge, UK
  (2009),
  \href{https://doi.org/10.1017/CBO9780511813924}{10.1017/CBO9780511813924}.

\bibitem{Ford:1985qh}
L.H.~Ford and A.~Vilenkin, \emph{{Global Symmetry Breaking in Two-dimensional
  Flat Space-time and in De Sitter Space-time}},
  \href{https://doi.org/10.1103/PhysRevD.33.2833}{\emph{Phys. Rev. D}
  {\bfseries 33} (1986) 2833}.

\bibitem{Starobinsky:1994bd}
A.A.~Starobinsky and J.~Yokoyama, \emph{{Equilibrium state of a selfinteracting
  scalar field in the De Sitter background}},
  \href{https://doi.org/10.1103/PhysRevD.50.6357}{\emph{Phys. Rev. D}
  {\bfseries 50} (1994) 6357}
  [\href{https://arxiv.org/abs/astro-ph/9407016}{{\ttfamily
  astro-ph/9407016}}].

\bibitem{Lazzari:2013boa}
G.~Lazzari and T.~Prokopec, \emph{{Symmetry breaking in de Sitter: a stochastic
  effective theory approach}},
  \href{https://arxiv.org/abs/1304.0404}{{\ttfamily 1304.0404}}.

\bibitem{Serreau:2013eoa}
J.~Serreau, \emph{{Renormalization group flow and symmetry restoration in de
  Sitter space}},
  \href{https://doi.org/10.1016/j.physletb.2014.01.058}{\emph{Phys. Lett. B}
  {\bfseries 730} (2014) 271}
  [\href{https://arxiv.org/abs/1306.3846}{{\ttfamily 1306.3846}}].

\bibitem{Guilleux:2015pma}
M.~Guilleux and J.~Serreau, \emph{{Quantum scalar fields in de Sitter space
  from the nonperturbative renormalization group}},
  \href{https://doi.org/10.1103/PhysRevD.92.084010}{\emph{Phys. Rev. D}
  {\bfseries 92} (2015) 084010}
  [\href{https://arxiv.org/abs/1506.06183}{{\ttfamily 1506.06183}}].

\bibitem{Dolgov:1981nw}
A.D.~Dolgov, \emph{{Conformal Anomaly and the Production of Massless Particles
  by a Conformally Flat Metric}},
  {\emph{{\href{http://www.jetp.ac.ru/cgi-bin/e/index/r/81/2/p417?a=list}{Sov.
  Phys. JETP}}} {\bfseries 54} (1981) 223}.

\bibitem{Benevides:2018mwx}
A.~Benevides, A.~Dabholkar and T.~Kobayashi, \emph{{To $B$ or not to $B$:
  Primordial magnetic fields from Weyl anomaly}},
  \href{https://doi.org/10.1007/JHEP11(2018)039}{\emph{JHEP} {\bfseries 11}
  (2018) 039} [\href{https://arxiv.org/abs/1808.08237}{{\ttfamily
  1808.08237}}].

\bibitem{Chen:2016nrs}
X.~Chen, Y.~Wang and Z.-Z.~Xianyu, \emph{{Loop Corrections to Standard Model
  Fields in Inflation}},
  \href{https://doi.org/10.1007/JHEP08(2016)051}{\emph{JHEP} {\bfseries 08}
  (2016) 051} [\href{https://arxiv.org/abs/1604.07841}{{\ttfamily
  1604.07841}}].

\bibitem{Chen:2016uwp}
X.~Chen, Y.~Wang and Z.-Z.~Xianyu, \emph{{Standard Model Background of the
  Cosmological Collider}},
  \href{https://doi.org/10.1103/PhysRevLett.118.261302}{\emph{Phys. Rev. Lett.}
  {\bfseries 118} (2017) 261302}
  [\href{https://arxiv.org/abs/1610.06597}{{\ttfamily 1610.06597}}].

\bibitem{Chen:2016hrz}
X.~Chen, Y.~Wang and Z.-Z.~Xianyu, \emph{{Standard Model Mass Spectrum in
  Inflationary Universe}},
  \href{https://doi.org/10.1007/JHEP04(2017)058}{\emph{JHEP} {\bfseries 04}
  (2017) 058} [\href{https://arxiv.org/abs/1612.08122}{{\ttfamily
  1612.08122}}].

\bibitem{Maleknejad:2012fw}
A.~Maleknejad, M.M.~Sheikh-Jabbari and J.~Soda, \emph{{Gauge Fields and
  Inflation}}, \href{https://doi.org/10.1016/j.physrep.2013.03.003}{\emph{Phys.
  Rept.} {\bfseries 528} (2013) 161}
  [\href{https://arxiv.org/abs/1212.2921}{{\ttfamily 1212.2921}}].

\bibitem{Durrer:2013pga}
R.~Durrer and A.~Neronov, \emph{{Cosmological Magnetic Fields: Their
  Generation, Evolution and Observation}},
  \href{https://doi.org/10.1007/s00159-013-0062-7}{\emph{Astron. Astrophys.
  Rev.} {\bfseries 21} (2013) 62}
  [\href{https://arxiv.org/abs/1303.7121}{{\ttfamily 1303.7121}}].

\bibitem{Subramanian:2015lua}
K.~Subramanian, \emph{{The origin, evolution and signatures of primordial
  magnetic fields}},
  \href{https://doi.org/10.1088/0034-4885/79/7/076901}{\emph{Rept. Prog. Phys.}
  {\bfseries 79} (2016) 076901}
  [\href{https://arxiv.org/abs/1504.02311}{{\ttfamily 1504.02311}}].

\bibitem{Turner:1987bw}
M.S.~Turner and L.M.~Widrow, \emph{{Inflation Produced, Large Scale Magnetic
  Fields}}, \href{https://doi.org/10.1103/PhysRevD.37.2743}{\emph{Phys. Rev. D}
  {\bfseries 37} (1988) 2743}.

\bibitem{Giovannini:2000dj}
M.~Giovannini and M.E.~Shaposhnikov, \emph{{Primordial magnetic fields from
  inflation?}}, \href{https://doi.org/10.1103/PhysRevD.62.103512}{\emph{Phys.
  Rev. D} {\bfseries 62} (2000) 103512}
  [\href{https://arxiv.org/abs/hep-ph/0004269}{{\ttfamily hep-ph/0004269}}].

\bibitem{Calzetta:1997ku}
E.A.~Calzetta, A.~Kandus and F.D.~Mazzitelli, \emph{{Primordial magnetic fields
  induced by cosmological particle creation}},
  \href{https://doi.org/10.1103/PhysRevD.57.7139}{\emph{Phys. Rev. D}
  {\bfseries 57} (1998) 7139}
  [\href{https://arxiv.org/abs/astro-ph/9707220}{{\ttfamily
  astro-ph/9707220}}].

\bibitem{Kandus:1999st}
A.~Kandus, E.A.~Calzetta, F.D.~Mazzitelli and C.E.M.~Wagner,
  \emph{{Cosmological magnetic fields from gauge mediated supersymmetry
  breaking models}},
  \href{https://doi.org/10.1016/S0370-2693(99)01389-1}{\emph{Phys. Lett. B}
  {\bfseries 472} (2000) 287}
  [\href{https://arxiv.org/abs/hep-ph/9908524}{{\ttfamily hep-ph/9908524}}].

\bibitem{Davis:2000zp}
A.-C.~Davis, K.~Dimopoulos, T.~Prokopec and O.~Tornkvist, \emph{{Primordial
  spectrum of gauge fields from inflation}},
  \href{https://doi.org/10.1016/S0370-2693(01)00138-1}{\emph{Phys. Lett. B}
  {\bfseries 501} (2001) 165}
  [\href{https://arxiv.org/abs/astro-ph/0007214}{{\ttfamily
  astro-ph/0007214}}].

\bibitem{Dimopoulos:2001wx}
K.~Dimopoulos, T.~Prokopec, O.~Tornkvist and A.C.~Davis, \emph{{Natural
  magnetogenesis from inflation}},
  \href{https://doi.org/10.1103/PhysRevD.65.063505}{\emph{Phys. Rev. D}
  {\bfseries 65} (2002) 063505}
  [\href{https://arxiv.org/abs/astro-ph/0108093}{{\ttfamily
  astro-ph/0108093}}].

\bibitem{Prokopec:2002uw}
T.~Prokopec, O.~Tornkvist and R.P.~Woodard, \emph{{One loop vacuum polarization
  in a locally de Sitter background}},
  \href{https://doi.org/10.1016/S0003-4916(03)00004-6}{\emph{Annals Phys.}
  {\bfseries 303} (2003) 251}
  [\href{https://arxiv.org/abs/gr-qc/0205130}{{\ttfamily gr-qc/0205130}}].

\bibitem{Prokopec:2002jn}
T.~Prokopec, O.~Tornkvist and R.P.~Woodard, \emph{{Photon mass from
  inflation}}, \href{https://doi.org/10.1103/PhysRevLett.89.101301}{\emph{Phys.
  Rev. Lett.} {\bfseries 89} (2002) 101301}
  [\href{https://arxiv.org/abs/astro-ph/0205331}{{\ttfamily
  astro-ph/0205331}}].

\bibitem{Prokopec:2003bx}
T.~Prokopec and R.P.~Woodard, \emph{{Vacuum polarization and photon mass in
  inflation}}, \href{https://doi.org/10.1119/1.1596180}{\emph{Am. J. Phys.}
  {\bfseries 72} (2004) 60}
  [\href{https://arxiv.org/abs/astro-ph/0303358}{{\ttfamily
  astro-ph/0303358}}].

\bibitem{Prokopec:2003iu}
T.~Prokopec and R.P.~Woodard, \emph{{Dynamics of superhorizon photons during
  inflation with vacuum polarization}},
  \href{https://doi.org/10.1016/j.aop.2004.01.012}{\emph{Annals Phys.}
  {\bfseries 312} (2004) 1}
  [\href{https://arxiv.org/abs/gr-qc/0310056}{{\ttfamily gr-qc/0310056}}].

\bibitem{Prokopec:2003tm}
T.~Prokopec and E.~Puchwein, \emph{{Photon mass generation during inflation: de
  Sitter invariant case}},
  \href{https://doi.org/10.1088/1475-7516/2004/04/007}{\emph{JCAP} {\bfseries
  04} (2004) 007} [\href{https://arxiv.org/abs/astro-ph/0312274}{{\ttfamily
  astro-ph/0312274}}].

\bibitem{Prokopec:2004au}
T.~Prokopec and E.~Puchwein, \emph{{Nearly minimal magnetogenesis}},
  \href{https://doi.org/10.1103/PhysRevD.70.043004}{\emph{Phys. Rev. D}
  {\bfseries 70} (2004) 043004}
  [\href{https://arxiv.org/abs/astro-ph/0403335}{{\ttfamily
  astro-ph/0403335}}].

\bibitem{Popov:2017xut}
F.K.~Popov, \emph{{Debye mass in de Sitter space}},
  \href{https://doi.org/10.1007/JHEP06(2018)033}{\emph{JHEP} {\bfseries 06}
  (2018) 033} [\href{https://arxiv.org/abs/1711.11010}{{\ttfamily
  1711.11010}}].

\bibitem{Prokopec:2007ak}
T.~Prokopec, N.C.~Tsamis and R.P.~Woodard, \emph{{Stochastic Inflationary
  Scalar Electrodynamics}},
  \href{https://doi.org/10.1016/j.aop.2007.08.008}{\emph{Annals Phys.}
  {\bfseries 323} (2008) 1324}
  [\href{https://arxiv.org/abs/0707.0847}{{\ttfamily 0707.0847}}].

\bibitem{Starobinsky:1986fx}
A.A.~Starobinsky, \emph{{STOCHASTIC DE SITTER (INFLATIONARY) STAGE IN THE EARLY
  UNIVERSE}}, \href{https://doi.org/10.1007/3-540-16452-9_6}{\emph{Lect. Notes
  Phys.} {\bfseries 246} (1986) 107}.

\bibitem{Kahya:2005kj}
E.O.~Kahya and R.P.~Woodard, \emph{{Charged scalar self-mass during
  inflation}}, \href{https://doi.org/10.1103/PhysRevD.72.104001}{\emph{Phys.
  Rev. D} {\bfseries 72} (2005) 104001}
  [\href{https://arxiv.org/abs/gr-qc/0508015}{{\ttfamily gr-qc/0508015}}].

\bibitem{Kahya:2006ui}
E.O.~Kahya and R.P.~Woodard, \emph{{One Loop Corrected Mode Functions for SQED
  during Inflation}},
  \href{https://doi.org/10.1103/PhysRevD.74.084012}{\emph{Phys. Rev. D}
  {\bfseries 74} (2006) 084012}
  [\href{https://arxiv.org/abs/gr-qc/0608049}{{\ttfamily gr-qc/0608049}}].

\bibitem{Prokopec:2006ue}
T.~Prokopec, N.C.~Tsamis and R.P.~Woodard, \emph{{Two Loop Scalar Bilinears for
  Inflationary SQED}},
  \href{https://doi.org/10.1088/0264-9381/24/1/011}{\emph{Class. Quant. Grav.}
  {\bfseries 24} (2007) 201}
  [\href{https://arxiv.org/abs/gr-qc/0607094}{{\ttfamily gr-qc/0607094}}].

\bibitem{Prokopec:2008gw}
T.~Prokopec, N.C.~Tsamis and R.P.~Woodard, \emph{{Two loop stress-energy tensor
  for inflationary scalar electrodynamics}},
  \href{https://doi.org/10.1103/PhysRevD.78.043523}{\emph{Phys. Rev. D}
  {\bfseries 78} (2008) 043523}
  [\href{https://arxiv.org/abs/0802.3673}{{\ttfamily 0802.3673}}].

\bibitem{Chen:1995ena}
L.-Y.~Chen, N.~Goldenfeld and Y.~Oono, \emph{{The Renormalization group and
  singular perturbations: Multiple scales, boundary layers and reductive
  perturbation theory}},
  \href{https://doi.org/10.1103/PhysRevE.54.376}{\emph{Phys. Rev. E} {\bfseries
  54} (1996) 376} [\href{https://arxiv.org/abs/hep-th/9506161}{{\ttfamily
  hep-th/9506161}}].

\bibitem{Burgess:2009bs}
C.P.~Burgess, L.~Leblond, R.~Holman and S.~Shandera, \emph{{Super-Hubble de
  Sitter Fluctuations and the Dynamical RG}},
  \href{https://doi.org/10.1088/1475-7516/2010/03/033}{\emph{JCAP} {\bfseries
  03} (2010) 033} [\href{https://arxiv.org/abs/0912.1608}{{\ttfamily
  0912.1608}}].

\bibitem{Tsamis:2005hd}
N.C.~Tsamis and R.P.~Woodard, \emph{{Stochastic quantum gravitational
  inflation}},
  \href{https://doi.org/10.1016/j.nuclphysb.2005.06.031}{\emph{Nucl. Phys. B}
  {\bfseries 724} (2005) 295}
  [\href{https://arxiv.org/abs/gr-qc/0505115}{{\ttfamily gr-qc/0505115}}].

\bibitem{Woodard:2005cw}
R.P.~Woodard, \emph{{A Leading logarithm approximation for inflationary quantum
  field theory}},
  \href{https://doi.org/10.1016/j.nuclphysbps.2005.04.056}{\emph{Nucl. Phys. B
  Proc. Suppl.} {\bfseries 148} (2005) 108}
  [\href{https://arxiv.org/abs/astro-ph/0502556}{{\ttfamily
  astro-ph/0502556}}].

\bibitem{Garbrecht:2013coa}
B.~Garbrecht, G.~Rigopoulos and Y.~Zhu, \emph{{Infrared correlations in de
  Sitter space: Field theoretic versus stochastic approach}},
  \href{https://doi.org/10.1103/PhysRevD.89.063506}{\emph{Phys. Rev. D}
  {\bfseries 89} (2014) 063506}
  [\href{https://arxiv.org/abs/1310.0367}{{\ttfamily 1310.0367}}].

\bibitem{Garbrecht:2014dca}
B.~Garbrecht, F.~Gautier, G.~Rigopoulos and Y.~Zhu, \emph{{Feynman Diagrams for
  Stochastic Inflation and Quantum Field Theory in de Sitter Space}},
  \href{https://doi.org/10.1103/PhysRevD.91.063520}{\emph{Phys. Rev. D}
  {\bfseries 91} (2015) 063520}
  [\href{https://arxiv.org/abs/1412.4893}{{\ttfamily 1412.4893}}].

\bibitem{Nacir:2016fzi}
D.~L\'opez~Nacir, F.D.~Mazzitelli and L.G.~Trombetta, \emph{{$O(N)$ model in
  Euclidean de Sitter space: beyond the leading infrared approximation}},
  \href{https://doi.org/10.1007/JHEP09(2016)117}{\emph{JHEP} {\bfseries 09}
  (2016) 117} [\href{https://arxiv.org/abs/1606.03481}{{\ttfamily
  1606.03481}}].

\bibitem{Janssen:2009pb}
T.M.~Janssen, S.P.~Miao, T.~Prokopec and R.P.~Woodard, \emph{{The Hubble
  Effective Potential}},
  \href{https://doi.org/10.1088/1475-7516/2009/05/003}{\emph{JCAP} {\bfseries
  05} (2009) 003} [\href{https://arxiv.org/abs/0904.1151}{{\ttfamily
  0904.1151}}].

\bibitem{Glavan:2020zne}
D.~Glavan, A.~Marunovi\'c, T.~Prokopec and Z.~Zahraee, \emph{{Abelian Higgs
  model in power-law inflation: the propagators in the unitary gauge}},
  \href{https://doi.org/10.1007/JHEP09(2020)165}{\emph{JHEP} {\bfseries 09}
  (2020) 165} [\href{https://arxiv.org/abs/2005.05435}{{\ttfamily
  2005.05435}}].

\bibitem{Berges:2004yj}
J.~Berges, \emph{{Introduction to nonequilibrium quantum field theory}},
  \href{https://doi.org/10.1063/1.1843591}{\emph{AIP Conf. Proc.} {\bfseries
  739} (2004) 3} [\href{https://arxiv.org/abs/hep-ph/0409233}{{\ttfamily
  hep-ph/0409233}}].

\bibitem{NeqQFT}
D.~Glavan and T.~Prokopec, \emph{{A pedestrian introduction to non-equilibrium
  QFT}},
  {\emph{\href{https://webspace.science.uu.nl/~proko101/LecturenotesNonEquilQFT.pdf}{\tt
  https://webspace.science.uu.nl/{\textasciitilde{}}proko101/LecturenotesNonEquilQFT.pdf}}
  }.

\bibitem{Kaya:2018qbj}
A.~Kaya, \emph{{Superhorizon Electromagnetic Field Background from Higgs Loops
  in Inflation}},
  \href{https://doi.org/10.1088/1475-7516/2018/03/046}{\emph{JCAP} {\bfseries
  03} (2018) 046} [\href{https://arxiv.org/abs/1801.02032}{{\ttfamily
  1801.02032}}].

\bibitem{Kobayashi:2019uqs}
T.~Kobayashi and M.S.~Sloth, \emph{{Early Cosmological Evolution of Primordial
  Electromagnetic Fields}},
  \href{https://doi.org/10.1103/PhysRevD.100.023524}{\emph{Phys. Rev. D}
  {\bfseries 100} (2019) 023524}
  [\href{https://arxiv.org/abs/1903.02561}{{\ttfamily 1903.02561}}].

\bibitem{DOnofrio:2012qeh}
M.~D'Onofrio, R.N.~Lerner and A.~Rajantie, \emph{{Electrically charged
  curvaton}}, \href{https://doi.org/10.1088/1475-7516/2012/10/004}{\emph{JCAP}
  {\bfseries 10} (2012) 004} [\href{https://arxiv.org/abs/1207.1063}{{\ttfamily
  1207.1063}}].

\bibitem{Goolsby-Cole:2015chd}
C.~Goolsby-Cole and L.~Sorbo, \emph{{On the electric charge of the observable
  Universe}},  \href{https://arxiv.org/abs/1511.07465}{{\ttfamily 1511.07465}}.

\bibitem{Janssen:2008px}
T.M.~Janssen, S.P.~Miao, T.~Prokopec and R.P.~Woodard, \emph{{Infrared
  Propagator Corrections for Constant Deceleration}},
  \href{https://doi.org/10.1088/0264-9381/25/24/245013}{\emph{Class. Quant.
  Grav.} {\bfseries 25} (2008) 245013}
  [\href{https://arxiv.org/abs/0808.2449}{{\ttfamily 0808.2449}}].

\bibitem{Chernikov:1968zm}
N.A.~Chernikov and E.A.~Tagirov, \emph{{Quantum theory of scalar fields in de
  Sitter space-time}},
  {\emph{{\href{http://www.numdam.org/item/AIHPA_1968__9_2_109_0/}{Ann. Inst.
  H. Poincare Phys. Theor. A}}} {\bfseries 9} (1968) 109}.

\bibitem{Bunch:1978yq}
T.S.~Bunch and P.C.W.~Davies, \emph{{Quantum Field Theory in de Sitter Space:
  Renormalization by Point Splitting}},
  \href{https://doi.org/10.1098/rspa.1978.0060}{\emph{Proc. Roy. Soc. Lond. A}
  {\bfseries 360} (1978) 117}.

\bibitem{Ford:1977in}
L.H.~Ford and L.~Parker, \emph{{Infrared Divergences in a Class of
  Robertson-Walker Universes}},
  \href{https://doi.org/10.1103/PhysRevD.16.245}{\emph{Phys. Rev. D} {\bfseries
  16} (1977) 245}.

\bibitem{Janssen:2009nz}
T.M.~Janssen and T.~Prokopec, \emph{{Regulating the infrared by mode matching:
  A Massless scalar in expanding spaces with constant deceleration}},
  \href{https://doi.org/10.1103/PhysRevD.83.084035}{\emph{Phys. Rev. D}
  {\bfseries 83} (2011) 084035}
  [\href{https://arxiv.org/abs/0906.0666}{{\ttfamily 0906.0666}}].

\bibitem{Lochan:2018pzs}
K.~Lochan, K.~Rajeev, A.~Vikram and T.~Padmanabhan, \emph{{Quantum correlators
  in Friedmann spacetimes: The omnipresent de Sitter spacetime and the
  invariant vacuum noise}},
  \href{https://doi.org/10.1103/PhysRevD.98.105015}{\emph{Phys. Rev. D}
  {\bfseries 98} (2018) 105015}
  [\href{https://arxiv.org/abs/1805.08800}{{\ttfamily 1805.08800}}].

\bibitem{Kobayashi:2014zza}
T.~Kobayashi and N.~Afshordi, \emph{{Schwinger Effect in 4D de Sitter Space and
  Constraints on Magnetogenesis in the Early Universe}},
  \href{https://doi.org/10.1007/JHEP10(2014)166}{\emph{JHEP} {\bfseries 10}
  (2014) 166} [\href{https://arxiv.org/abs/1408.4141}{{\ttfamily 1408.4141}}].

\bibitem{Frob:2014zka}
M.B.~Fr\"ob, J.~Garriga, S.~Kanno, M.~Sasaki, J.~Soda, T.~Tanaka et~al.,
  \emph{{Schwinger effect in de Sitter space}},
  \href{https://doi.org/10.1088/1475-7516/2014/04/009}{\emph{JCAP} {\bfseries
  04} (2014) 009} [\href{https://arxiv.org/abs/1401.4137}{{\ttfamily
  1401.4137}}].

\bibitem{Banyeres:2018aax}
M.~Banyeres, G.~Dom\`enech and J.~Garriga, \emph{{Vacuum birefringence and the
  Schwinger effect in (3+1) de Sitter}},
  \href{https://doi.org/10.1088/1475-7516/2018/10/023}{\emph{JCAP} {\bfseries
  10} (2018) 023} [\href{https://arxiv.org/abs/1809.08977}{{\ttfamily
  1809.08977}}].

\bibitem{Moss:2016uix}
I.~Moss and G.~Rigopoulos, \emph{{Effective long wavelength scalar dynamics in
  de Sitter}}, \href{https://doi.org/10.1088/1475-7516/2017/05/009}{\emph{JCAP}
  {\bfseries 05} (2017) 009}
  [\href{https://arxiv.org/abs/1611.07589}{{\ttfamily 1611.07589}}].

\bibitem{Prokopec:2017vxx}
T.~Prokopec and G.~Rigopoulos, \emph{{Functional renormalization group for
  stochastic inflation}},
  \href{https://doi.org/10.1088/1475-7516/2018/08/013}{\emph{JCAP} {\bfseries
  08} (2018) 013} [\href{https://arxiv.org/abs/1710.07333}{{\ttfamily
  1710.07333}}].

\end{thebibliography}\endgroup

\end{document}